\documentclass[]{IEEEtran}

\usepackage{amsfonts}
\usepackage{tabu}
\usepackage{booktabs}
\usepackage{float}
\usepackage{gensymb}
\usepackage{textcomp}
\usepackage{dblfloatfix}
\usepackage{amsmath}
\usepackage{algorithm}
\usepackage{algpseudocode}
\usepackage{pifont}
\usepackage{verbatim}
\usepackage{graphicx}
\usepackage{seqsplit}
\usepackage{caption}
\usepackage{enumitem}
\usepackage{subcaption}
\usepackage{url}
\usepackage{gensymb}
\usepackage{array}
\usepackage{multirow}

\DeclareMathOperator*{\argmin}{arg\,min}

\begin{document}

\title{Light Ears: Information Leakage via Smart Lights}

\author{\IEEEauthorblockN{Anindya Maiti \textit{and} Murtuza Jadliwala}\\
\IEEEauthorblockA{University of Texas at San Antonio\\
Email: a.maiti@ieee.org, murtuza.jadliwala@utsa.edu}}

\maketitle

\begin{abstract}
Modern Internet-enabled smart lights promise energy efficiency and many additional capabilities over traditional lamps. However, these connected lights create a new attack surface, which can be maliciously used to violate users' privacy and security. In this paper, we design and evaluate novel attacks that take advantage of light emitted by modern smart bulbs in order to infer users' private data and preferences. The first two attacks are designed to infer users' audio and video playback by a systematic observation and analysis of the multimedia-visualization functionality of smart light bulbs. The third attack utilizes the infrared capabilities of such smart light bulbs to create a covert-channel, which can be used as a gateway to exfiltrate user's private data out of their secured home or office network. A comprehensive evaluation of these attacks in various real-life settings confirms their feasibility and affirms the need for new privacy protection mechanisms.
\end{abstract}

\section{Introduction}
\label{sec:intro}


Smart lighting products, marketed as energy-efficient replacements of traditional incandescent and fluorescent lamps with several novel functionalities enabled by their Internet connectivity, have soared in popularity in recent years \cite{statista20142017,statista20162023} 
A common feature among most smart lights is the ability to remotely control the lights, over a Wi-Fi, Bluetooth, or Zigbee network. Many of the current generation smart lights are also LED-based, which enables fine-grained customization of color and intensity of the light being emitted from these bulbs. Few advanced smart lights are also equipped with infrared capabilities, intended to aid surveillance cameras in low visibility environments.

Lighting products have traditionally not been an attractive target of security/privacy-related threats because conventional lamps typically do not have access to sensitive user information. However, as modern smart lights are usually connected to users' home or office network (either directly or via a communication hub) and can be controlled using users' mobile devices, they are poised to become a much more attractive target for security/privacy attacks than before. \emph{In this paper, we investigate how modern smart lights can be exploited to infer users' private information.}

In this direction, we first focus on exploiting (details in Sections \ref{sec:audio-visualization} and \ref{sec:video-visualization}) a new feature of modern smart lights, known as \textit{multimedia-visualization}. Multimedia-visualization is intended for use in conjunction with a song or video playing on a nearby media player, which results in a vibrant lighting effect that is synchronized with the tones present in the audio or the dominant colors in the video stream, respectively. While such immersive audio-visual or ambient lighting effects can be entertaining and relaxing, we speculate that it can also lead to loss of privacy if not properly safeguarded. Consider a scenario where an curious adversary can observe the changing light intensities/colors of a multimedia-visualizing smart light installed inside a user's residence (likely through a window). \textit{Can the adversary determine what song/video the user is playing, only by analyzing the changing light intensities/colors of the smart light?} If yes, then these attacks can have significant privacy implications for smart light users. For instance, the US Video Privacy Protection Act (1988) was enacted to prevent abuse of users' media consumption information, which can potentially reveal fine-grained personal interests and preferences. \emph{Our first goal in this paper is to comprehensively evaluate the feasibility of such attacks that passively employ outputs from smart lights to infer users' personal media (both, audio and video) consumption. }

Next, we comprehensively study and evaluate the feasibility (details in Section \ref{sec:infrared-attack}) of exploiting a smart light's infrared lighting functionality to invisibly exfiltrate a user's private data out of his/her secured personal device or network. We show that such an attack can be accomplished by carefully manipulating and controlling (possible on modern smart lights) the infrared light to create a ``covert-channel'' between the smart light and an adversary with infrared sensing capability. With the help of a malicious agent on the user's smartphone or computer, the adversary can encode private information residing on these devices and then later transmit it over the infrared covert-channel residing on the smart light. 
Moreover, as several popular brands of smart lights do not require any form of authorization for controlling lights (infrared or otherwise) on the local network, any application installed on the target user's smartphone or computer can safely act as the malicious data exfiltration agent. 
The overarching goal of this paper is to highlight the vulnerable state of personal information of smart light users, outline system design parameters that lead to these vulnerabilities and discuss potential protection strategies against such threats.


\section{Technical Background}
\label{sec:background}
LIFX \cite{lifx} and Phillips Hue \cite{hue} are two of the most popular\linebreak commercially-available smart light systems. 
These smart lighting systems support millions of colors and multiple shades of white, have fine-grained brightness and saturation controls, possess wireless communication capabilities and can be controlled (wirelessly) using software platforms and mobile apps developed by the manufacturers.
Several novel applications have been developed for such smart lighting systems by leveraging on their unique wireless communication and fine-grained control capabilities. For instance, both LIFX and Phillips Hue bulbs support \emph{multimedia-visualization} by means of the manufacturer-provided mobile app or via third-party apps such as Light DJ \cite{lightdj}, lifxDynamic/hueDynamic \cite{huedynamic}, etc.

The multimedia-visualization feature of the smart lighting system, which can be turned on using the system's mobile app, is used to change or modulate the brightness and/or color of user-selected bulb(s) of the system in real-time. During \textbf{\textit{audio-visualization}}, the change in brightness is made to reflect either the surrounding sound levels captured using an on-device microphone (more ambient noise), or the direct output of an audio playing application (less or no ambient noise). During \textbf{\textit{video-visualization}}, the change in color and brightness is made to reflect the dominant color and brightness level in the current frame of the video being played. The mobile app remotely controls the bulb by periodically sending it specially formatted packets. A careful analysis of the local network traffic shows that audio-visualizing applications transmit approximately 10 packets to the bulb per second, whereas video-visualizing applications transmit approximately 1 packet per second.
Communication in case of the LIFX bulbs happens via an 802.11 access point, whereas the Phillips Hue bulb employs 802.15.4 (Zigbee) protocol to communicate with the mobile app. 
Each packet in the LIFX protocol \cite{lifxlan} consists of three headers followed by a payload:

\begin{itemize}[leftmargin=0.2in]
\item
The headers -- \textit{frame header}, \textit{frame address} and \textit{protocol header} -- includes details on how to process the frame and its payload, and also specifies the destination bulb.
\item
The \textit{payload} describes the desired hue, brightness and saturation levels.
\end{itemize}

The Phillips Hue platform has a very similar communication protocol, but employs an additional hub between the 802.11 access point and the bulb. The hub translates TCP/IP packets in to Zigbee Light Link \cite{zigbeelightlink} packets comprehensible to the Hue bulbs. Upon receiving a packet, the target bulb parses the packet and changes its color, brightness and saturation based on the payload information. When multimedia-visualization (audio or video) is enabled, a stream of packets from the controlling mobile application creates a real-time visual effect with the output light.

\begin{figure}[]
\centering
\includegraphics[width= 0.99\linewidth]{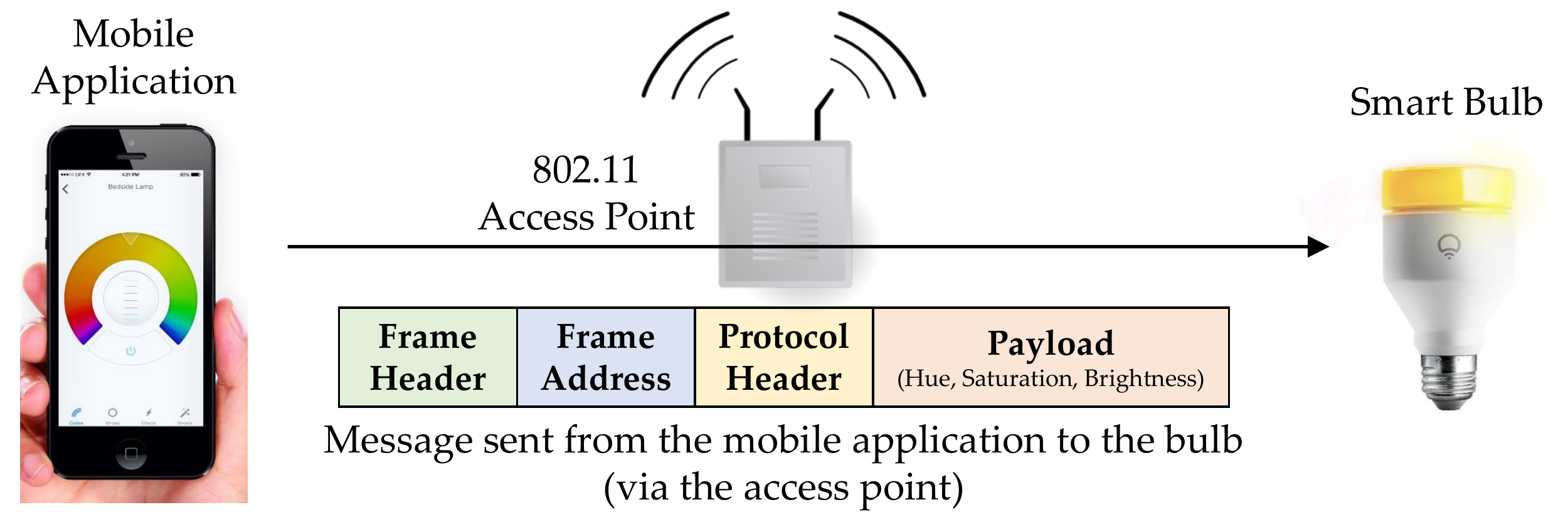}
\caption{The LIFX protocol, where a mobile application can control the bulb using specially formatted packets.}
\label{background-packets}
\end{figure}

\noindent
\textbf{Color models.} Next, we briefly discuss the physical properties of light emitted from multi-color LED bulbs. These properties are utilized in the attacks presented later in the paper. Most colors observed on multi-color LED bulbs (such as the LIFX and Phillips Hue), are not the same as the electromagnetic VIBGYOR spectrum. Each color in the VIBGYOR spectrum has a unique wavelength, whereas the colors `seen' on most multi-color LED bulbs are made of three additive primary colors. Almost all visible colors can be constructed for the human eye by the additive color mixing of three colors that are in widely spaced regions of the visible spectrum. If the three colors of light can be mixed to produce white, 
they are called primary colors and the de facto additive primary colors are Red, Green and Blue (RGB). It is important to note that three-primary-color models (such as RGB) work only to `trick' the human eye, which perceives color using three types of cone cells (S, M, and L) \cite{roorda1999arrangement}. To illustrate with an example, the violet color seen on a multi-color LED bulb is a combination of red ($\lambda$ = 620-750 $nm$), green ($\lambda$ = 495-570 $nm$) and blue ($\lambda$ = 450-495 $nm$) lights in $\langle 1:0:2\rangle$ ratio, whereas violet light in a natural rainbow has a wavelength $\lambda$ = 380-450 $nm$. $RGB\langle 1:0:2\rangle$ happens to excite the S, M, and L cone cells in the same way a 380-450 $nm$ wavelength light does. Therefore, we humans perceive $RGB\langle 1:0:2\rangle $ as violet. Moreover. the RGB model enables the production of several other perceptual colors, such as the color magenta or $RGB\langle 1:0:1\rangle $, which does not have its own place in the electromagnetic spectrum. A RGB color $\langle r,g,b\rangle$, where $r,g,b\in{[0,1]}$, also self-defines the lightness (or darkness) of the color. In relation to multi-color LED bulbs, at $RGB\langle 1,1,0\rangle$ the bulb outputs yellow light at maximum brightness, whereas at $RGB\langle 0.1,0.1,0\rangle$ the bulb outputs yellow light at roughly $10\%$ of maximum brightness.

\begin{figure}[]
\centering
\begin{subfigure}{0.37\linewidth}
\centering
\includegraphics[width=\textwidth]{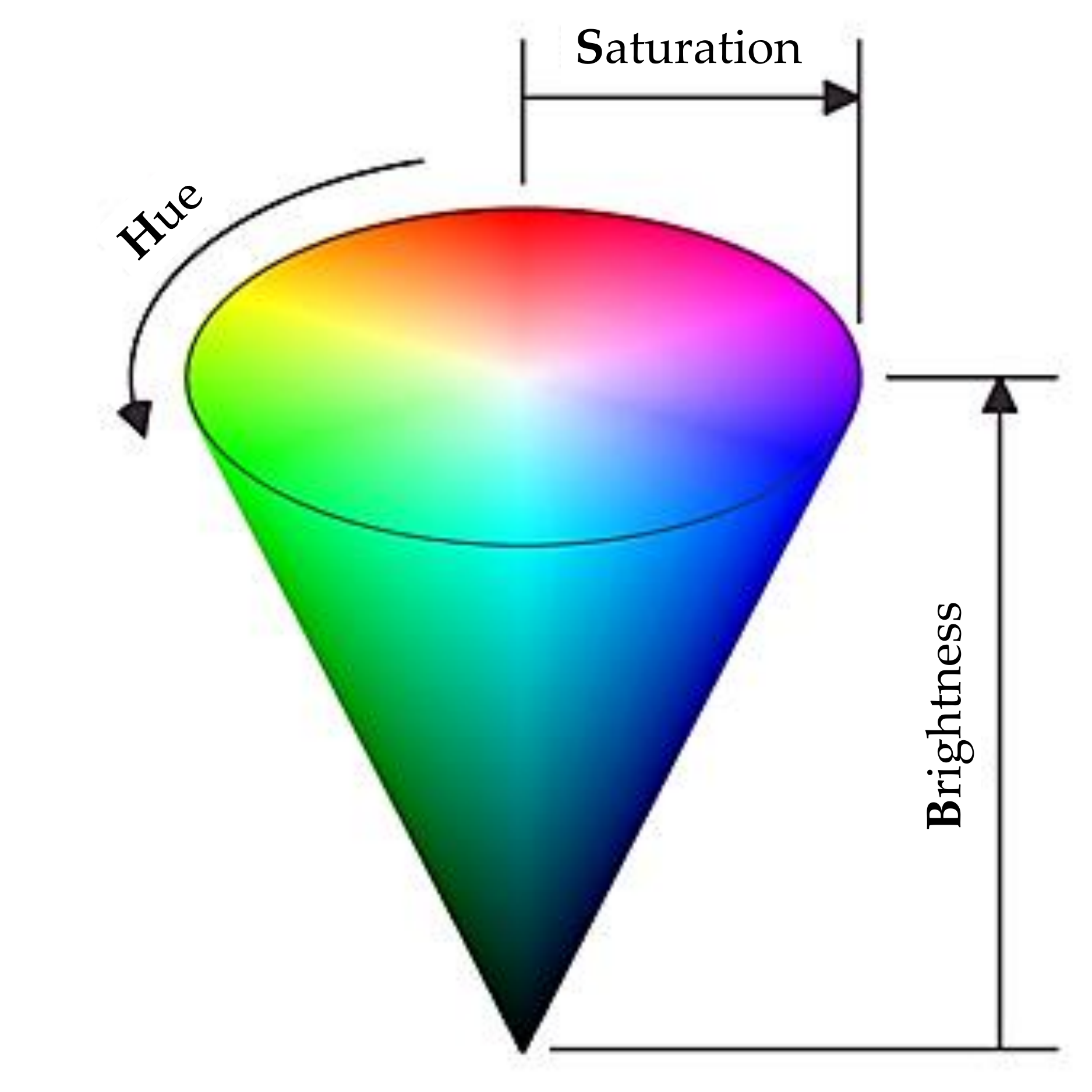}
\caption{}
\label{HSB-cycle}
\end{subfigure}
\hfill
\begin{subfigure}{0.51\linewidth}
\centering
\includegraphics[width=\textwidth]{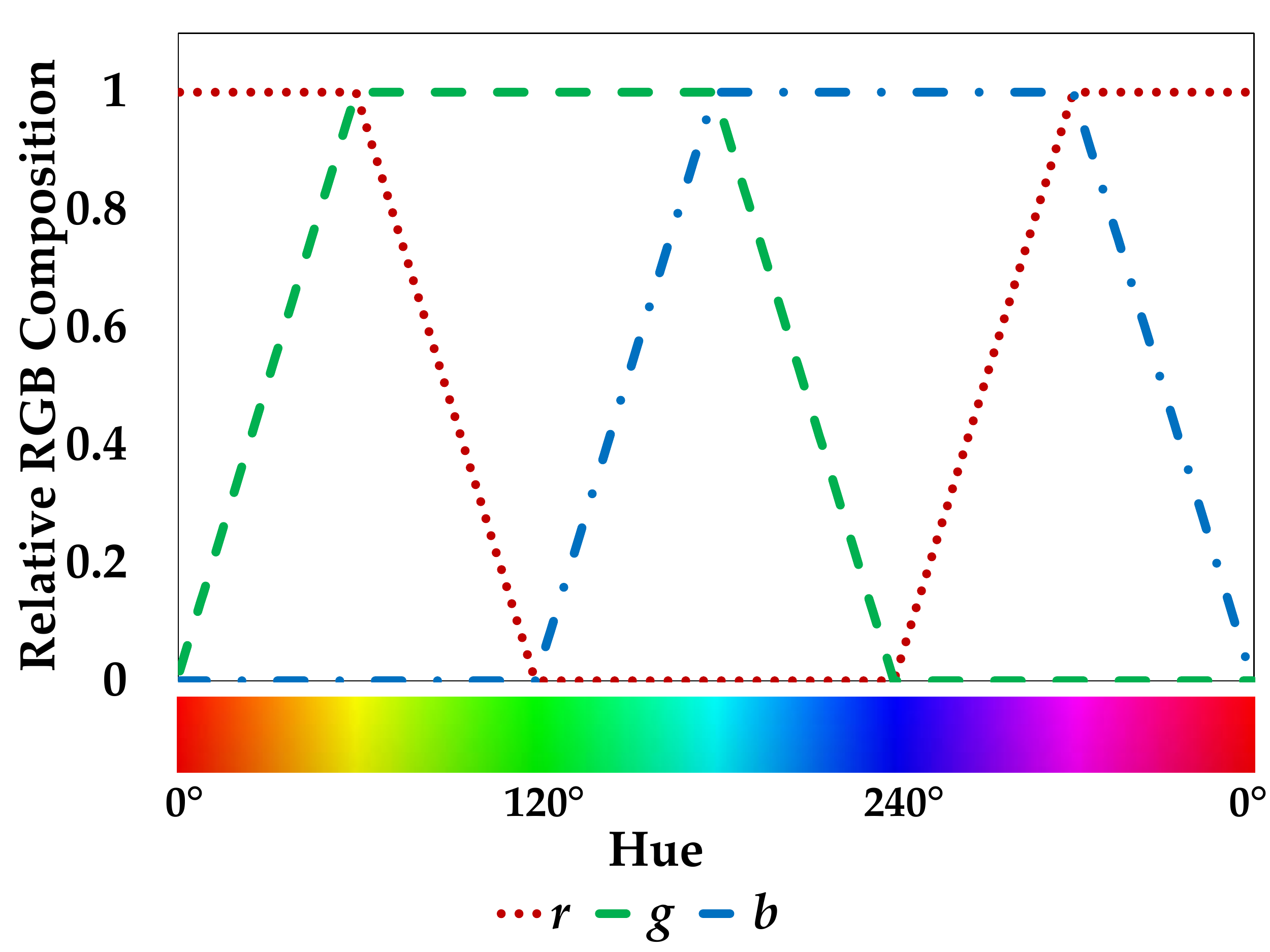}
\caption{}
\label{HSB-cycle-composition}
\end{subfigure}
\caption{(a) -- The HSB cone; (b) -- RGB composition of HSB colors, at full saturation.}
\label{background-color3-4}
\end{figure}

An alternate representation of RGB colors is commonly done using the HSB (hue, saturation and brightness) model (Figure \ref{HSB-cycle}). 
The \textit{hue} represents all colors (achievable by mixing RGB) in 360 degrees, with $0\degree$, $120\degree$  and $240\degree$ being Red, Green and Blue, respectively. At full \textit{saturation} of compound colors, only two of the three primary colors are mixed (Figure \ref{HSB-cycle-composition}), which is represented by the outer surface of the HSB cone. Adding the third primary color reduces saturation and the saturation level moves closer to the center of the HSB cone. When all three primary colors are mixed in equal proportion, it results in white (or zero saturation). In HSB model, \textit{brightness} is defined as the intensity of the dominant primary color(s). However, the perceived brightness (or relative luminance) depends on several biological properties of the eye and physical properties of electromagnetic radiation \cite{jameson1964theory,rossi1999neural}. 

In summary, RGB can be translated to HSB using Equation \ref{hsbeq}, as follows:
{\small
\begin{equation}
\begin{aligned}
H=\cos ^{-1}\left({\frac {{\frac {1}{2}}((r-g)+(r-b))}{(r-g)^{2}+(r-b)(g-b)}}\right)^{\frac {1}{2}};\\
S=1\,-\,{\frac {3}{(r+g+b)}}\,\min(r,g,b);\\
B=\operatorname {max} (r,g,b);\\
r,g,b\in{[0,1]}.
\end{aligned}
\label{hsbeq}
\end{equation}
}

\noindent
\textbf{Infrared lighting.} In addition to RGB colors, some smart lights are also equipped with infrared capabilities. For example, the LIFX+ series of bulbs support 950 $nm$ infrared lighting, which is invisible to the human eye, but useful for security cameras without built-in night-vision lighting. The brightness of the infrared light on LIFX+ bulbs can be controlled using packets with a payload describing the power level. A power level of zero indicates that the infrared LEDs will not be used, while a power level of 65535 indicates that the infrared channel should be set to the maximum possible value.

\section{Adversary Model}
\label{sec:adversarymodel}
For the private information inference threats that take advantage of the audio-visualizing and video-visualizing functionalities of smart lights, we assume a passive adversary whose goal is to infer a target user's media consumption by visually eavesdropping on the smart bulb's output (i.e., light emitted by bulb), without actively attacking the user's wireless (Wi-Fi, Bluetooth, or Zigbee) network or appliances. The user's wireless network is assumed to be secured against eavesdropping attacks, for example using WPA2, so the adversary cannot perform direct analysis of the packets sent to the smart bulb (from the controlling mobile application). In addition to visual eavesdropping instruments, such as light and color sensors, the adversary also needs sufficient computational and storage resources to initially create a comprehensive reference library of media items (songs and videos), and then match eavesdropped light patterns to items in the reference library.

For the data exfiltration threat using infrared-enabled smart lights, we assume an adversary whose goal is to exfiltrate data out of a target user's network or personal device, which is again assumed to be secured against wireless eavesdropping attacks. For the data exfiltration attack to work, the adversary has to additionally install a malicious software agent on the target user's device, for example, their smartphone or computer, that connects to the same network as their infrared-capable smart bulbs. This can be achieved by social engineering attacks \cite{krombholz2015advanced}, or by tricking the user in to installing a Trojan application \cite{felt2011survey}. The malicious software agent is responsible for encoding target user's private data (accessible on-device or on the network) in a format suitable for infrared communication, and transmits the encoded data using the user's infrared-enabled smart light. 
We also assume that the malicious software agent cannot directly communicate with an adversarial server over Internet (for example, due to a firewall), or that the user's network is air-gapped. In addition to infrared sensing hardware, for successfully executing this attack the adversary also needs sufficient computational and storage resources to record time-series of infrared light and process them to reconstruct the intended private data. In this paper, we explain and evaluate the feasibility of the above threats by means of a representative smart lighting system, namely, the LIFX system. We also assess the generalizability of our results by evaluating the proposed threats on an alternate system such as Phillips Hue.

\begin{figure*}[h]
\centering
\begin{subfigure}{\linewidth}
\centering
\includegraphics[width=\textwidth]{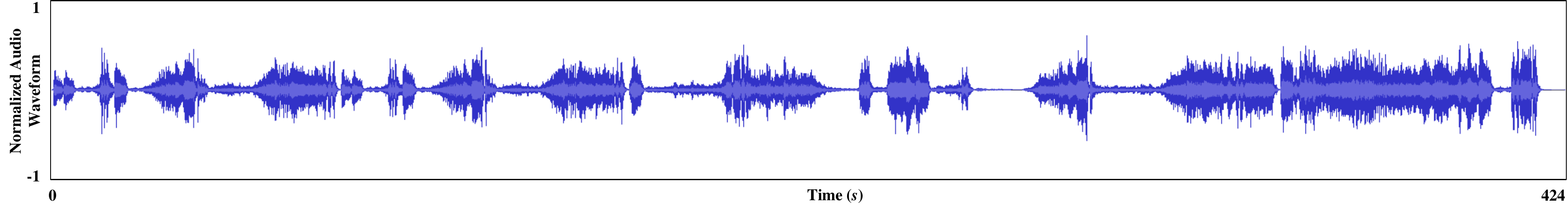}
\caption{}
\label{B5a}
\end{subfigure}
\hfill
\begin{subfigure}{\linewidth}
\centering
\includegraphics[width=\textwidth]{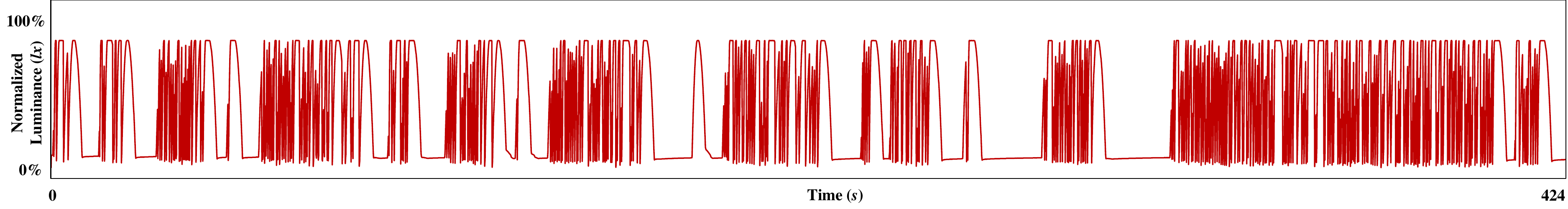}
\caption{}
\label{B5b}
\end{subfigure}
\caption{(a) Audio waveform of \textit{Beethoven's 5th Symphony} (duration of 7 minutes and 4 seconds); (b) Fluctuations in brightness of an audio-visualizing smart bulb output (in static hue mode), as captured by a BH1751FVI luminance meter.}
\label{B5}
\end{figure*}

\section{Audio Inference Threat}
\label{sec:audio-visualization}
We first outline the working of audio-visualizing smart lights, followed by the proposed inference attack on audio-visualizing lights. 
\textit{In the audio-visualization mode, the smart bulb reacts to the high and low tones present in the input audio stream by fluctuating its output light brightness}. Smart light mobile applications typically offer two different light coloring modes during audio-visualization. In \textbf{\textit{static hue}} mode, the bulb hue does not change unless manually changed, while in the \textbf{\textit{random hue}} mode the bulb periodically changes to a random hue at full saturation. In both modes, the input audio stream affects only the brightness level of the bulb, and the bulb hue (static or random) has no correlation to the audio.

\subsection{Effect of Sound on Bulb Brightness}
\label{audio-effect}
A simple observation with naked eyes and ears already gave us the intuition that a smart bulb's brightness fluctuates more with higher audio amplitudes. To precisely study this audio-visualization property of smart light bulbs, we conducted a systematic analysis of several raw audio tracks by precisely measuring the corresponding fluctuations in a LIFX bulb's brightness levels. For this we created an exploratory setup (details in Appendix B), where we played a few sample audio tracks multiple times in static hue mode, and observed the corresponding time-series of the bulb's brightness profile measured using a luminance meter. 
Our observations from this exploratory setup resulted in two significant conclusions, both of which are instrumental in the design of our attack framework. First, the observed luminance ($\delta_l$) of the bulb fluctuates when a relatively high amplitude is detected in the audio stream. These high amplitudes may represent various types of tones, such as high vocals or drum beats, which varies on the type of song being played. The luminance fluctuation subsides when the audio amplitude is low. This phenomenon is shown with an example in Figure \ref{B5}. It is evident from this example that there exists a clear correlation between the audio waveform of the song (say $A_1)$ and its corresponding ``luminance-profile'' (say $L_{A_1}$). In other words, a luminance-profile is the time-series of brightness values of a audio-visualizing bulb when the song is playing.

Our second observation is that for a given song, the luminance-profile suffers minor distortions across multiple recordings. In other words, if the luminance-profile of a song $A_1$ captured during two different playbacks is denoted by $L_{A_1}'$ and $L_{A_1}''$, respectively, then $L_{A_1}'$ and $L_{A_1}''$ will be similar but not identical. These minor distortions can be attributed to factors such as varying network latencies, packet loss due to network congestion and ambient audio noise. Also, because the brightness level is updated only every 100 milliseconds (10 $Hz$), an imperfect alignment of the song's starting point within the 100 milliseconds interval can cause a slightly different luminance-profile that follows. That being said, the similarity between $L_{A_1}'$ and $L_{A_1}''$ is generally very good. We utilize this property to design our audio inference framework which is based on \textit{elastic} time-series matching (Section \ref{audio-matching}) and is immune to such minor distortions.

Let us further clarify the notion of `brightness' in these luminance-profiles captured by the luminance sensor.
The total amount of visible light coming out from a source (the bulb in our case) is described using the term \textit{luminous flux}, which is the luminous energy released per unit time by the source and is measured in SI units of lumen or $lm$.
\textit{Illuminance}, on the other hand, is the amount of luminous flux incident on a surface, and is measured in lux or $lx$ ($lx=lm/m^2$). The (normalized) brightness values present in the luminance-profile time series that we construct (Figure \ref{B5b}) are nothing but the amount of luminous energy incident on the BH1751FVI photodiode (luminance meter used in our setup) divided by surface area of the sensor. In other words, the brightness values in the luminance-profile time-series are nothing but the illuminance values captured by the photodiode or luminance sensor.
It can be difficult to accurately measure illuminance using a photodiode that is also sensitive to infrared and/or ultraviolet radiations. However, as the BH1751FVI's spectral sensitivity is very close to the human eye (Figure \ref{background-color1}), the sensitivity of the photodiode to infrared and ultraviolet radiations is not significant. 

\subsection{Factors Affecting Song Identification}
\label{audio-matching}
Next, let's discuss how we can use a luminance-profile in \textit{static} and \textit{random} hue modes to uniquely identify songs. In static hue mode the relative change in bulb's illuminance will cause proportional gain or drop on the luminance meter. This is true for all hues, including compound hues such as yellow or cyan, because the spectral sensitivity remains constant when the hue is not changing. As a result, the normalized luminance-profile of a song will be similar across different colors, as long as the hue is not changed between the start and end of a song playback. This allows us to create a reference library of luminance-profiles of songs using a single hue, and use it to match songs across any hue. 

\begin{figure}[]
\centering
\begin{subfigure}{0.49\linewidth}
\centering
\includegraphics[width=\textwidth]{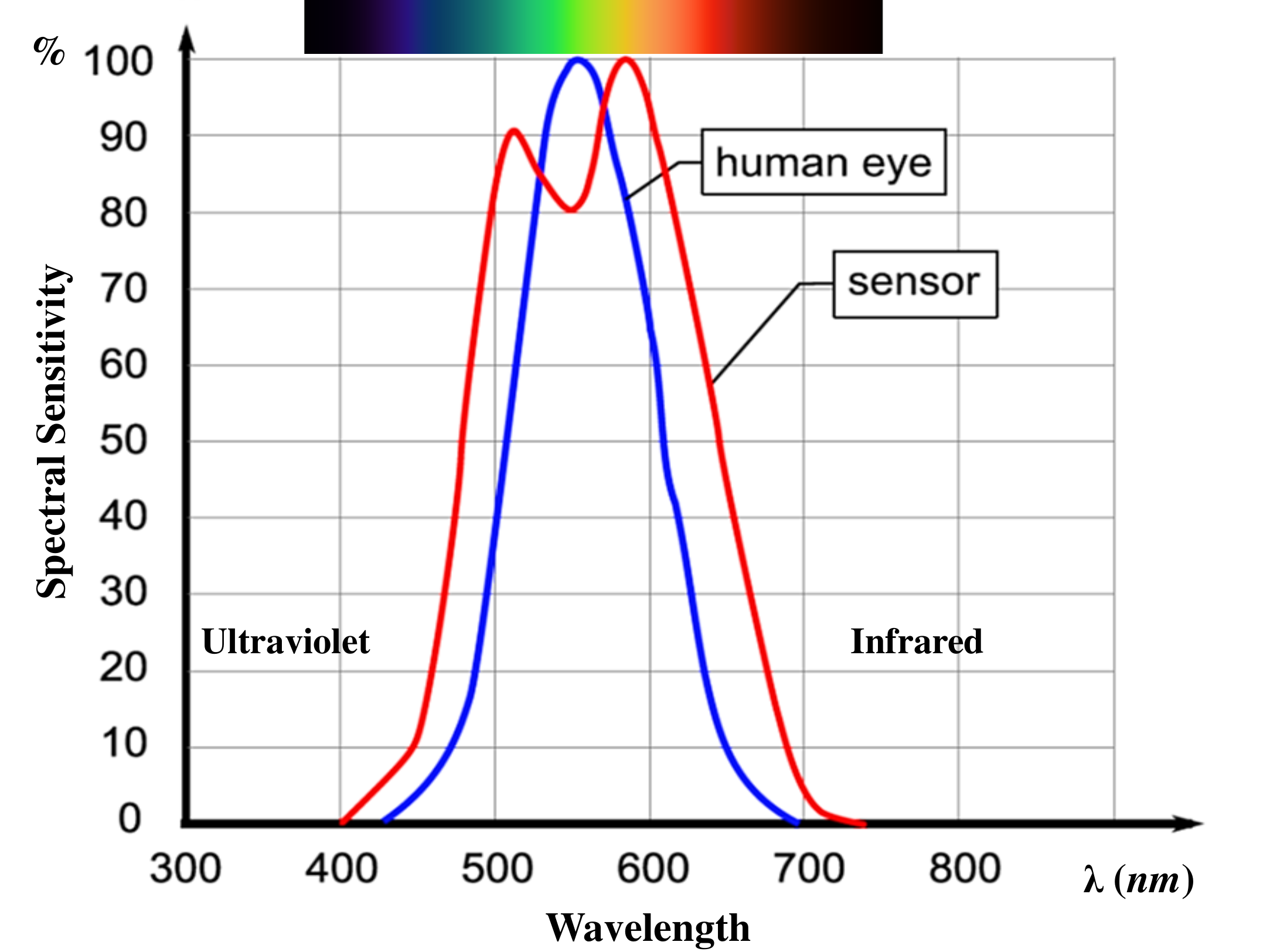}
\caption{}
\label{background-color1}
\end{subfigure}
\hfill
\begin{subfigure}{0.49\linewidth}
\centering
\includegraphics[width=\textwidth]{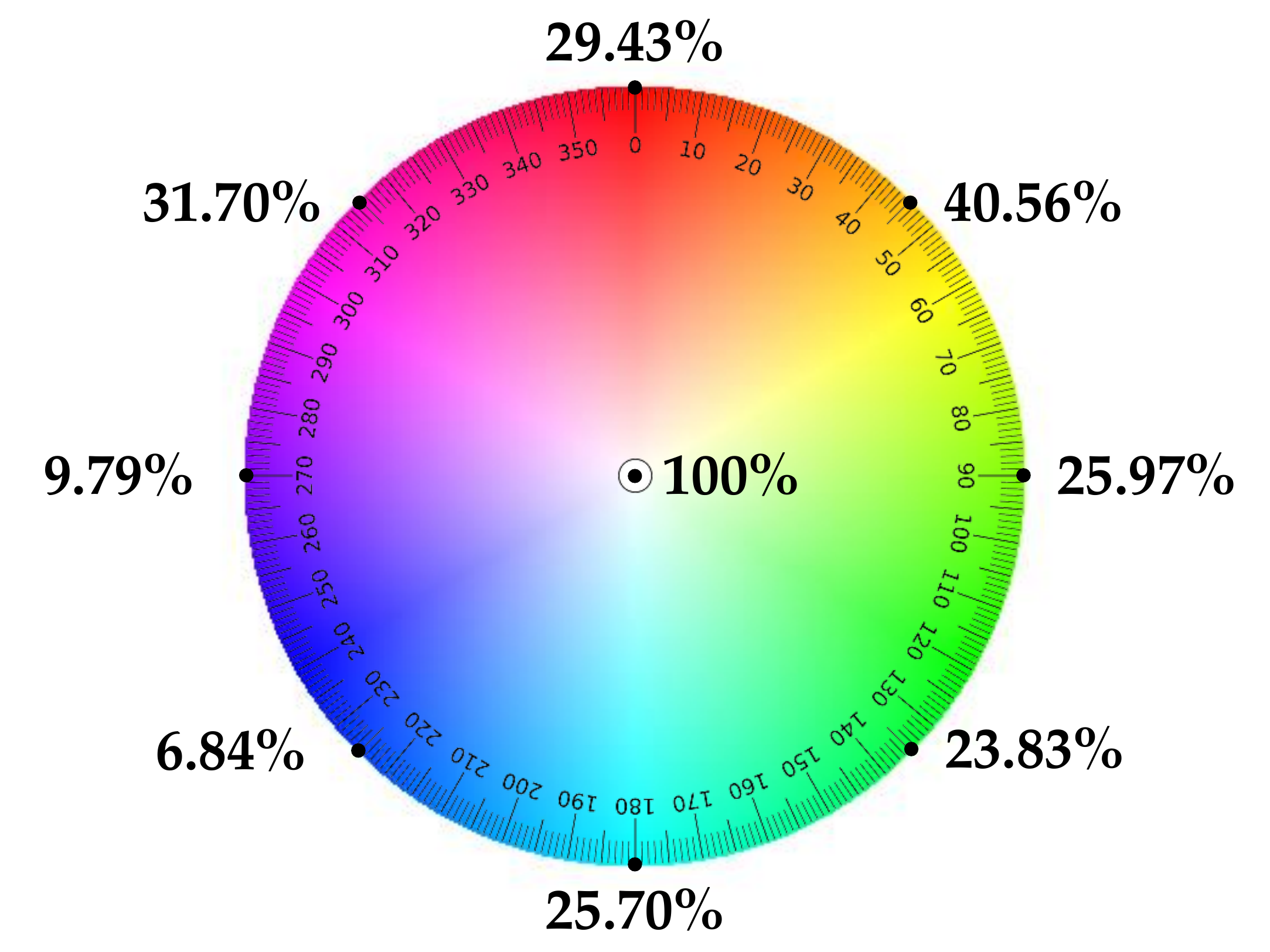}
\caption{}
\label{background-color2}
\end{subfigure}
\caption{(a) Luminance meter's response to the electromagnetic spectrum; (b) Luminance meter's response to different LIFX bulb hues (using RGB-HSB color model) at fixed brightness and full saturation.}
\label{background-color1-2}
\end{figure}

Matching luminance-profiles in random hue mode is, however, more complex. The luminance-profile in random hue mode is affected by two factors: the brightness (which is dependent on the song's amplitudes), and the luminance meter's sensitivity towards the hues that appear during the playback (which is independent of the song). From Figure \ref{background-color1} we already know that the luminance meter's sensitivity varies for different hues, which can create unwanted fluctuations in the luminance meter's response while recording the luminance-profile of a song (Figure \ref{background-color2}). This may result in incorrect song matching and poor inference performance. Moreover, characterizing the sensitivity of the BH1751FVI photodiode for each hue is not helpful because the RGB color spectrum does not have the same physical properties as the VIBGYOR spectrum (as explained in Section \ref{sec:background}). Also, certain perceptual colors do not appear on the VIBGYOR spectrum.

\begin{figure}[]
\centering
\includegraphics[width=\linewidth]{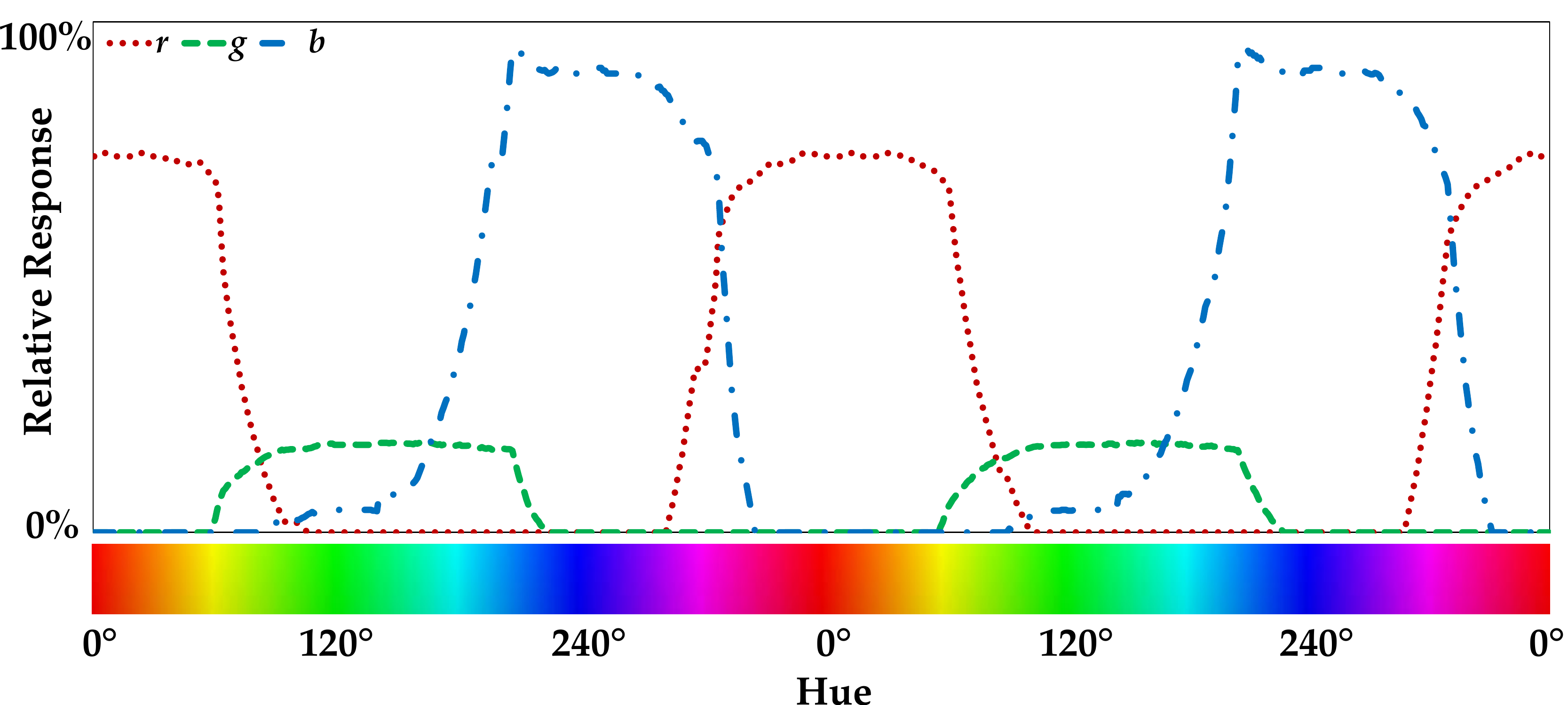}
\caption{Relative response of $\langle r,g,b\rangle$ components for the entire HSB hue cycle, at full brightness and saturation.}
\label{rgb-cd-cycle}
\end{figure}

A solution to this problem can be achieved by using a RGB sensor in conjunction with the luminance meter. A RGB sensor can help breakdown the composition of random hues coming out of the smart bulb, which can be used to uniquely identify the hues. Subsequently, the highest observed luminance level of each hue can be used to normalize the luminance values of all instances of corresponding hues, that appear during a song. Once normalized per hue, the resultant time-series becomes independent of changes in luminance due to changing hues, and only depends on the song amplitudes. This hue-independent luminance-profile should be similar across two different recordings, even when the order of hues is completely different. To test the feasibility of this approach, we conducted additional exploratory experiments using a Vernier GDX-LC RGB sensor. We cycled the bulb through the entire HSB hue palette at full brightness and saturation, and recorded the responses observed on the RGB sensor. Figure \ref{rgb-cd-cycle} shows that the observed response of Red ($\delta_r$), Green ($\delta_g$) and Blue ($\delta_b$) components have different characteristics. $\delta_b$ has the highest peak response, followed by $\delta_r$ and $\delta_g$. Relative luminance calibrations made by LED bulb manufacturers (so that at fixed brightness, red, green and blue lights are perceived by the human eye to be of the same level) play a significant factor in the differing peak response. As seen in Figure \ref{background-color1}, our eyes are more sensitive to green than red and blue. As a result, red and blue colors are calibrated to be more luminous than green, so as avoid a greenish-white perception for $RGB\langle 1:1:1\rangle$. We also observed that the color composition does not follow the ideal HSB model (Figure \ref{HSB-cycle-composition}), with some hues missing a primary color and some hues where all three primary colors are present (which should not happen at full saturation). This observation can be attributed to imperfect RGB sensing by the sensor \cite{rgbmanual}, and/or possible design limitations of the RGB LEDs present in the bulb. These observations were consistent with another bulb of same make and model, which reduces the possibility of having manufacturing defects in our particular bulb.

\begin{figure}[]
\centering
\begin{subfigure}{0.49\linewidth}
\centering
\includegraphics[width=\textwidth]{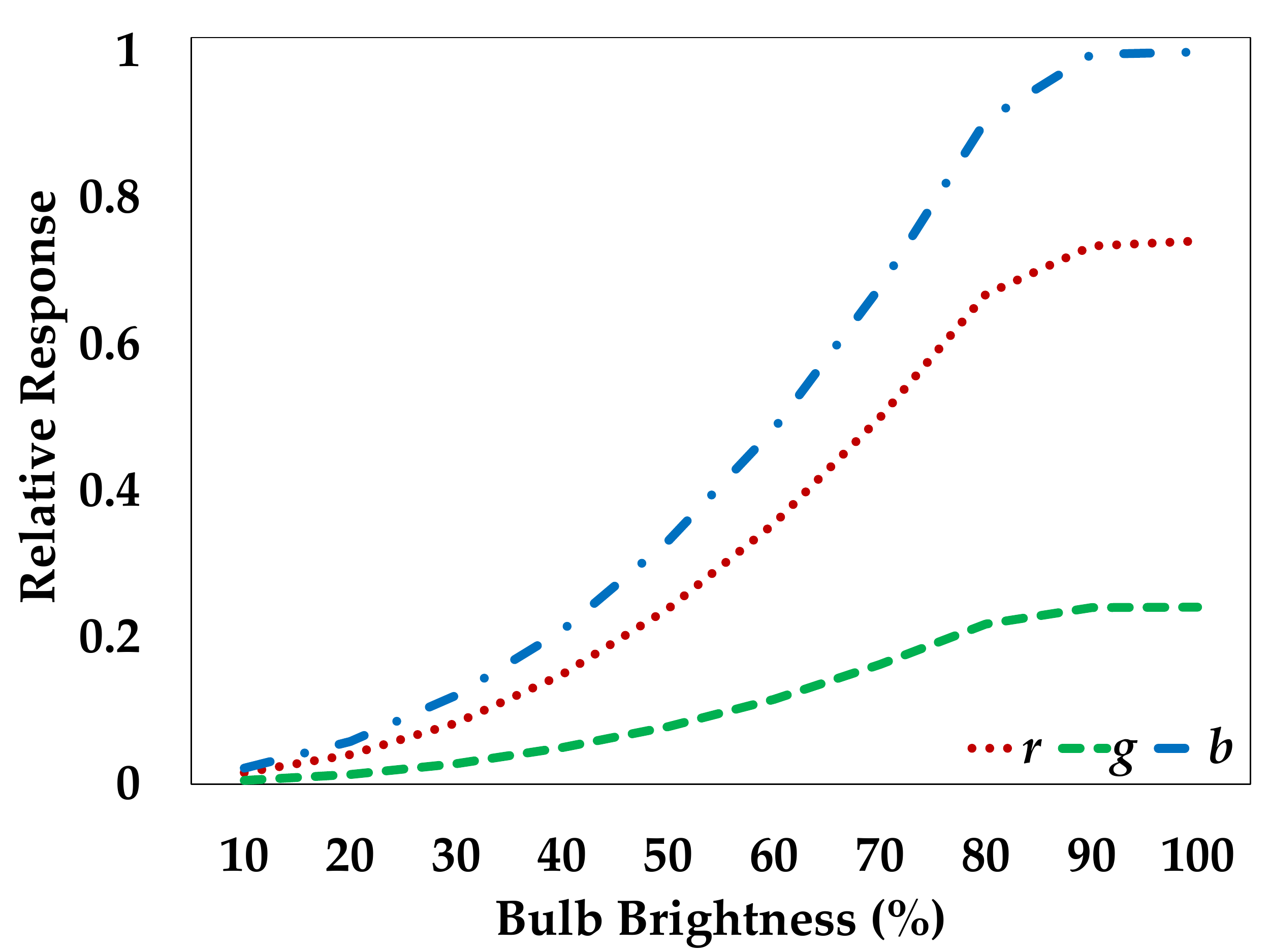}
\caption{}
\label{white-rgb}
\end{subfigure}
\hfill
\begin{subfigure}{0.49\linewidth}
\centering
\includegraphics[width=\textwidth]{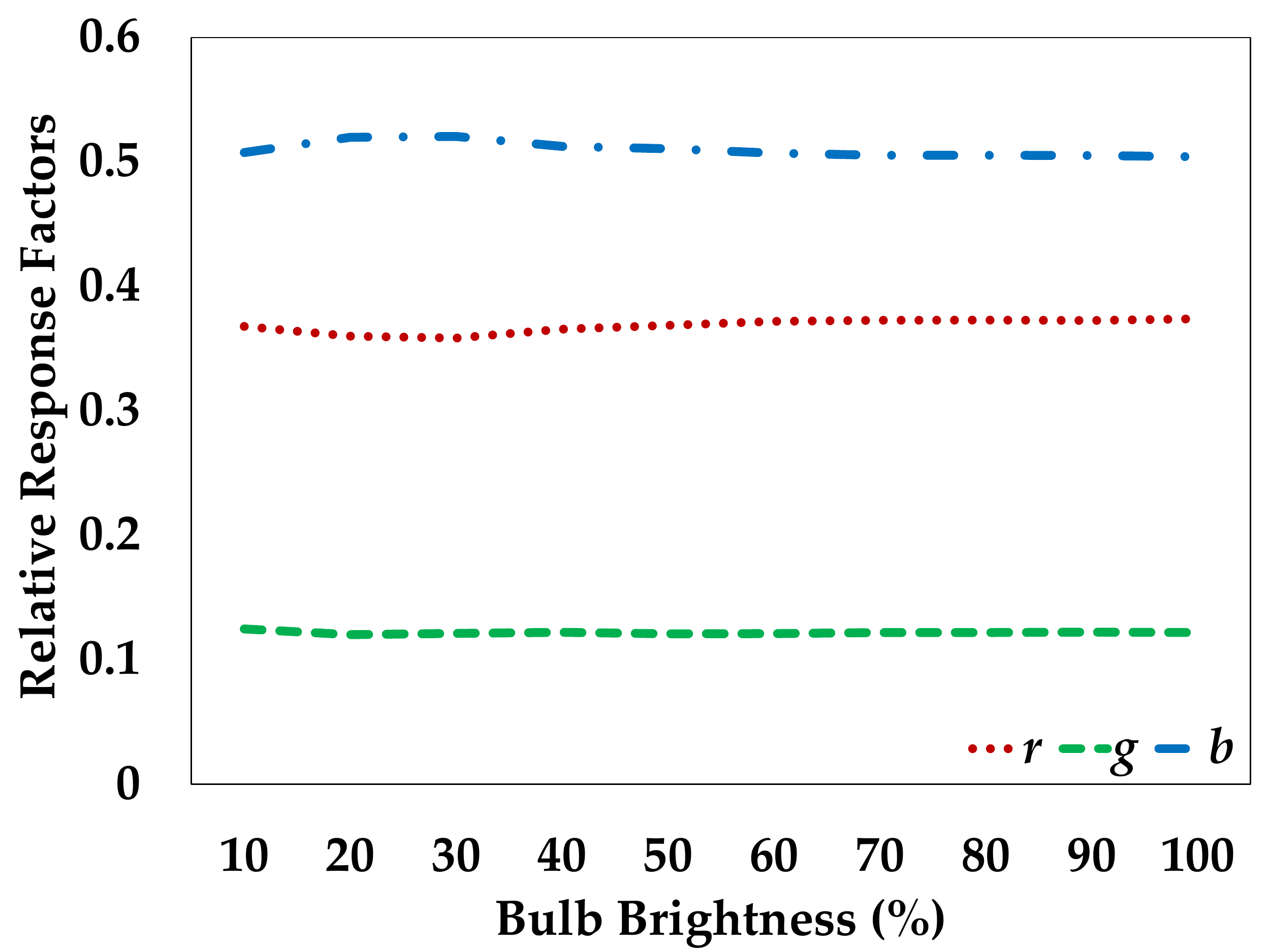}
\caption{}
\label{white-rgb-factors}
\end{subfigure}
\caption{(a) Observed response of $\langle r,g,b\rangle$ components $\langle \delta_r,\delta_g,\delta_b\rangle$ for $RGB\langle 1:1:1\rangle$, and (b) Response factors for $RGB\langle 1:1:1\rangle$ at different brightness levels.}
\label{white-rgb-and-factors}
\end{figure}

While the above observations are not ideal, we can still approximately identify individual hues based on their observed RGB response ratios $\langle \delta_r:\delta_g:\delta_b\rangle$. Certain hues neighboring to $0\degree$ and $240\degree$ cannot be uniquely identified because of only one primary color being present. However, that does not affect the proposed approach of normalization based on highest luminance of the identified hue, because hues neighboring to $0\degree$ and $240\degree$ are physically still red and blue lights, respectively. Moreover, the response of the only primary color remains roughly the same for these neighboring hues (Figure \ref{rgb-cd-cycle}), at fixed brightness. In order to consistently identify a light hue at different brightness levels based on its RGB composition, the response ratio $\langle \delta_r:\delta_g:\delta_b\rangle$ should also remain constant across varying brightness levels. To test if this holds true, we analyzed $RGB\langle 1:1:1\rangle$ at different brightness levels and recorded the corresponding $\delta_r$, $\delta_g$ and $\delta_b$ values observed on the GDX-LC sensor. Figure \ref{white-rgb} shows the increasing $\langle \delta_r$, $\delta_g$ and $\delta_b\rangle$ values as the brightness was increased, and Figure \ref{white-rgb-factors} shows the response factors. As the response factors \big($\frac {\delta_r}{(\delta_r+\delta_g+\delta_b)}$, $\frac {\delta_g}{(\delta_r+\delta_g+\delta_b)}$ and $\frac {\delta_b}{(\delta_r+\delta_g+\delta_b)}$\big) remain fairly constant across the entire brightness range, we can in fact use $\langle \delta_r:\delta_g:\delta_b\rangle$ to uniquely identify light hues at any brightness level. 

\begin{figure*}[]
\centering
\begin{subfigure}{0.61\linewidth}
\centering
\includegraphics[width=\linewidth]{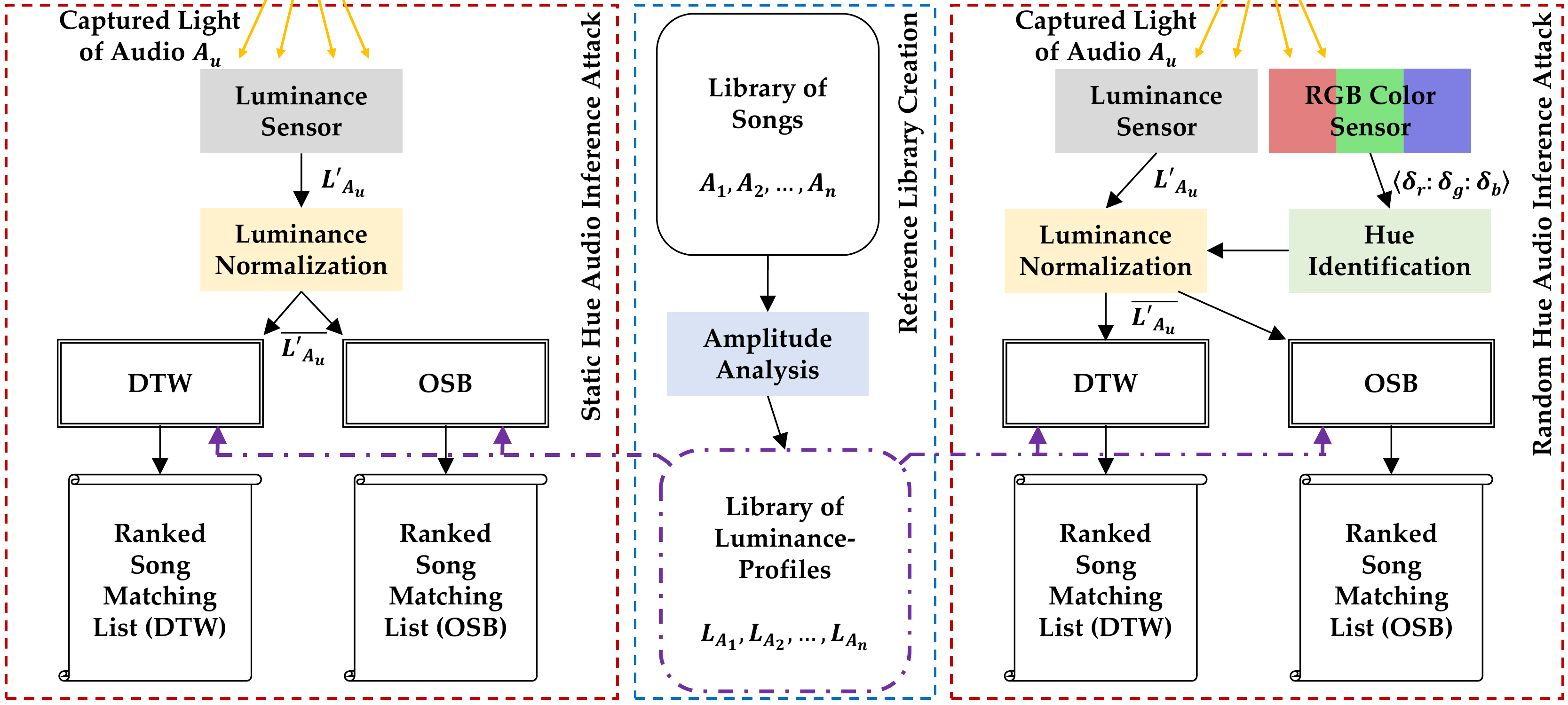}
\caption{}
\label{attackframework-a}
\end{subfigure}
\hfill
\begin{subfigure}{0.367\linewidth}
\centering
\includegraphics[width=\linewidth]{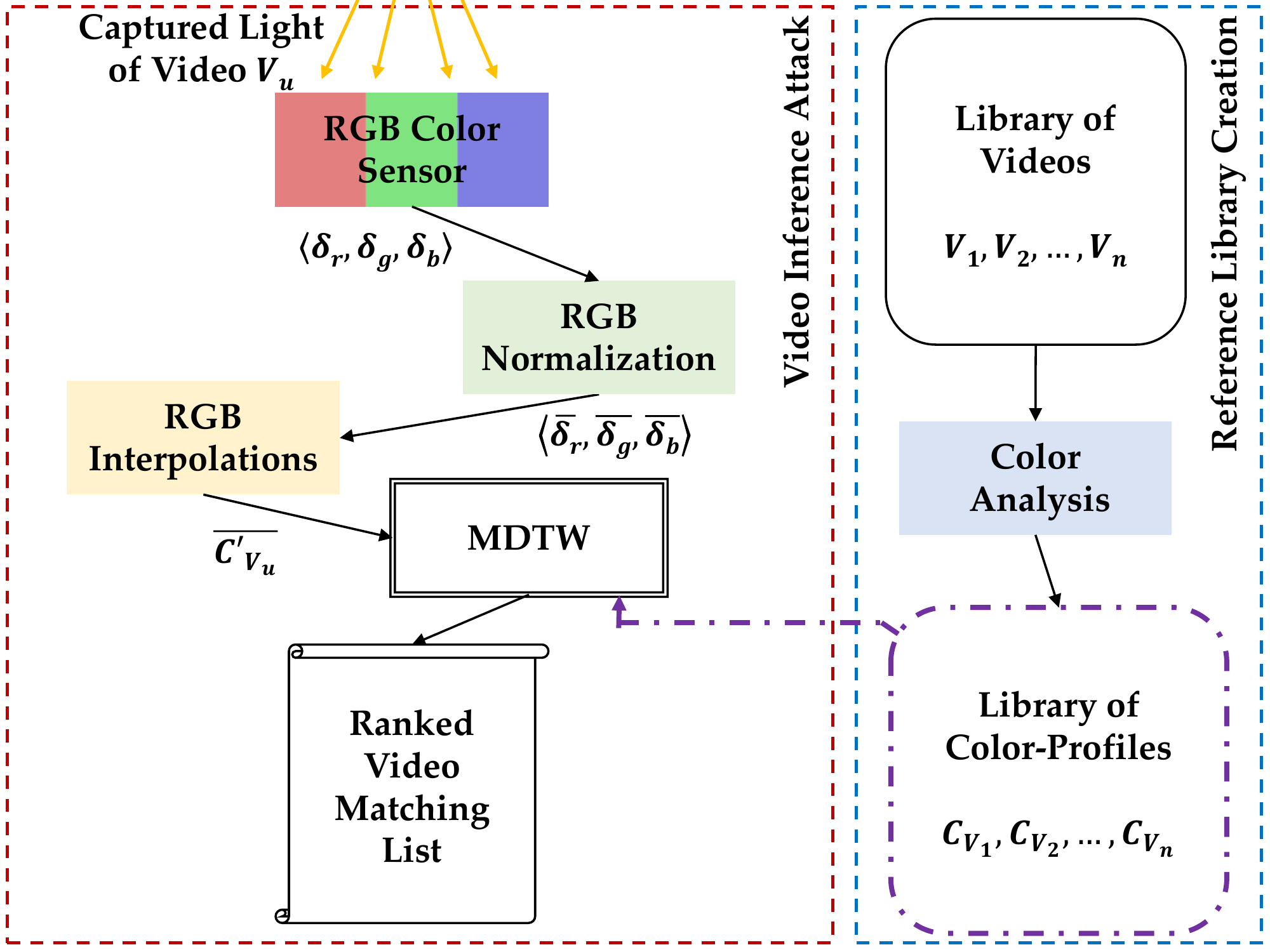}
\caption{}
\label{attackframework-v}
\end{subfigure}
\caption{(a) Audio Inference Framework; (b) Video Inference Framework.}
\label{attackframework-fig}
\end{figure*}

\begin{figure}[htp]
\centering
\begin{subfigure}{0.49\linewidth}
\centering
\includegraphics[width=\textwidth]{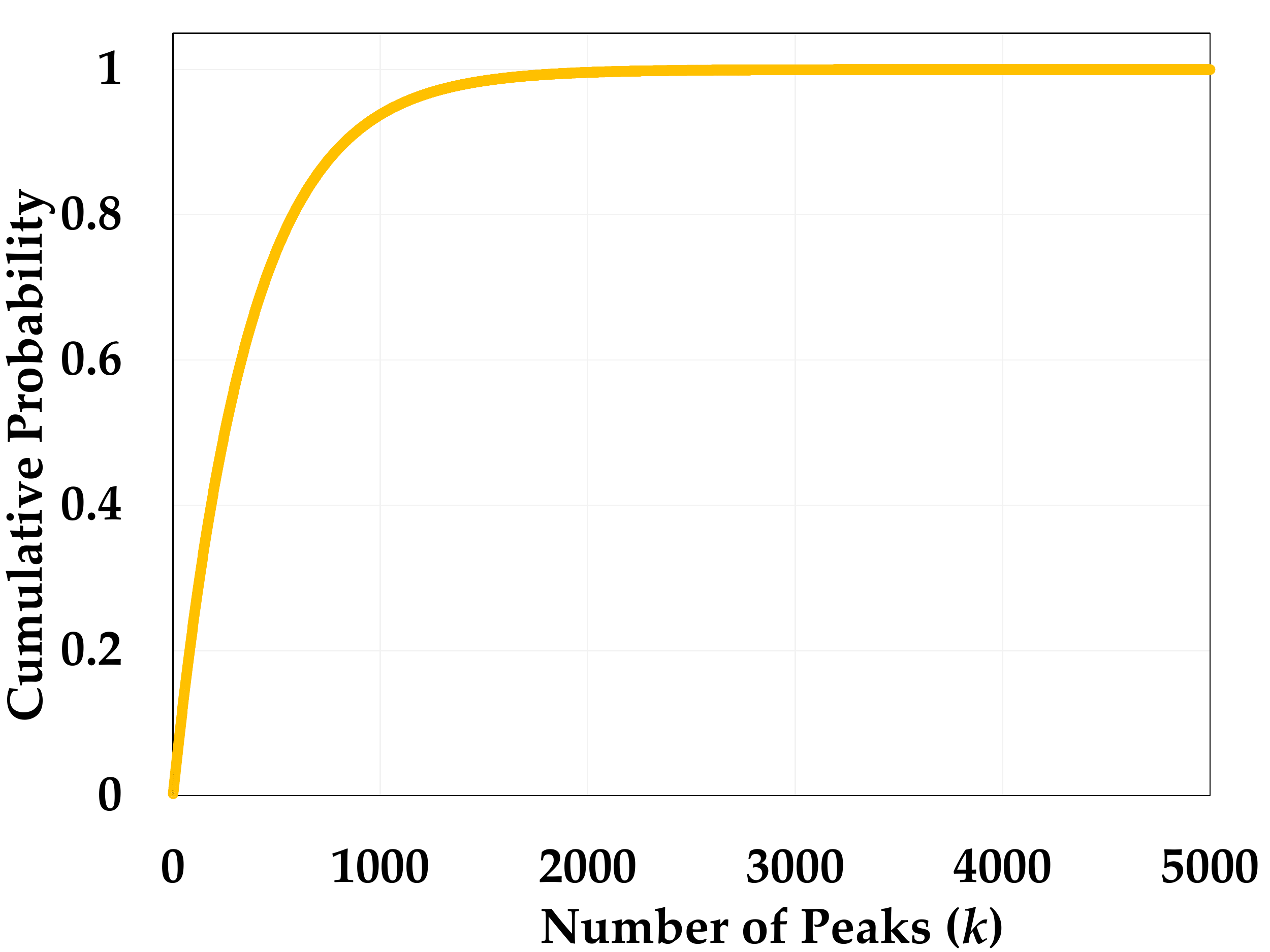}
\caption{}
\label{peaks-cdf}
\end{subfigure}
\hfill
\begin{subfigure}{0.49\linewidth}
\centering
\includegraphics[width=\textwidth]{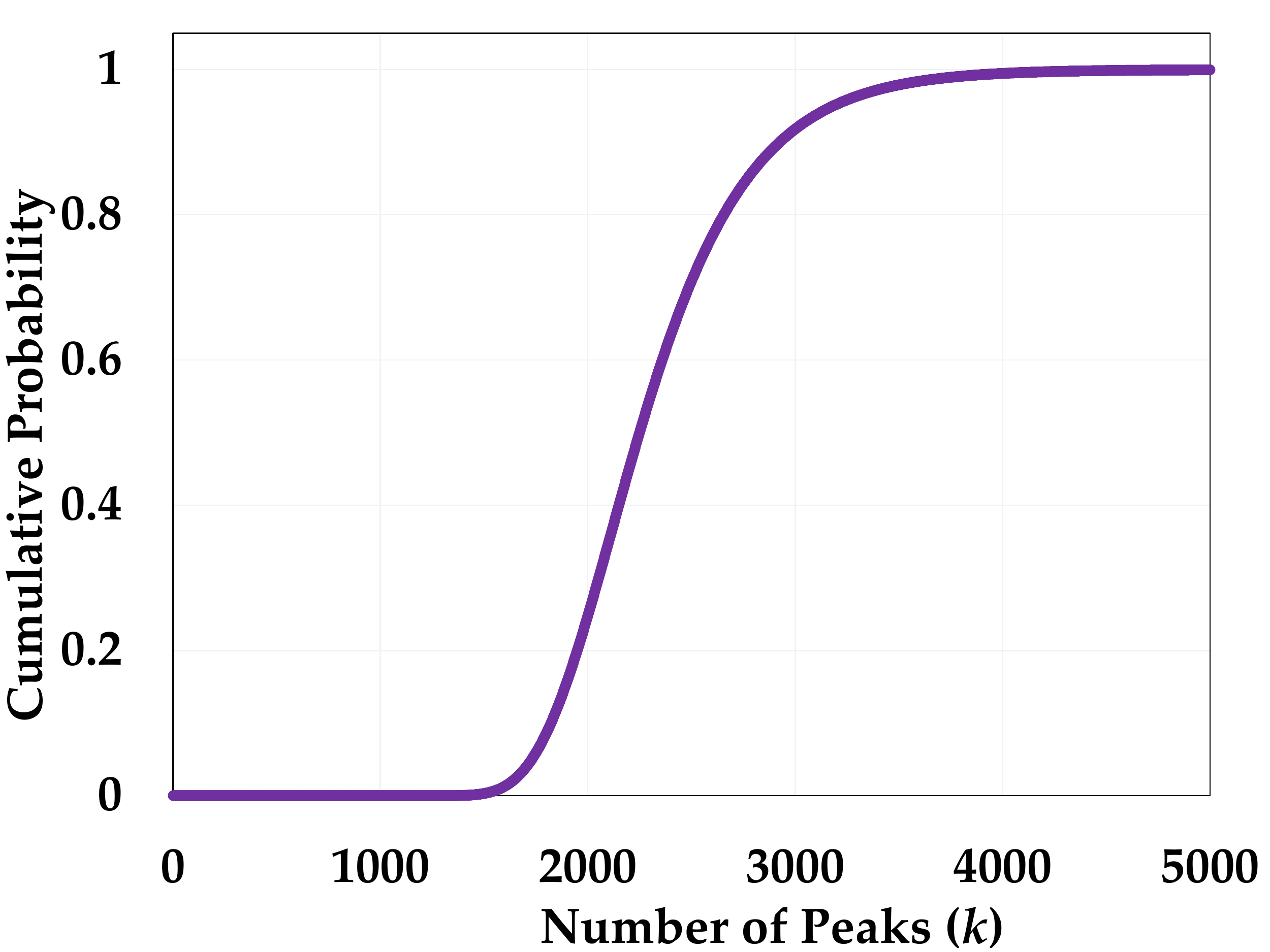}
\caption{}
\label{peaks-cdf-360}
\end{subfigure}
\caption{(a) Cumulative probability distribution of a specific hue appearing within first $k$ peaks; (b) Cumulative probability distribution of all 360 hues appearing within first $k$ peaks.}
\label{peaks-probability}
\end{figure}

The next challenge in normalizing a random hue luminance-profile is to ensure that we observe all hues at their highest luminance points at least once, based on which all occurrences of the corresponding hues will be normalized. The highest luminance point is generally observed at the \textit{peaks} in fluctuations that occur when a high amplitude is detected in the input audio. The number of peaks per minute varies significantly between different types of music. With a conservative assumption of 20 peaks per minutes, it will take a minimum of 18 minutes to learn the highest luminance values of 360 hues (one per degree). However, because the hue changes randomly in random hue mode, the probability that a specific hue appears on a peak is $\frac {1}{360}$ and the probability that a hue appears within first $k$ peaks can be represented as the cumulative probability of a geometric distribution: $1-(1-\frac {1}{360})^{k}$. The probability that all 360 hues appear at least once within first $k$ peaks can be modeled as a classical occupancy problem in probability theory \cite{gittelsohn1969occupancy}. Figure \ref{peaks-cdf} shows the cumulative probability distribution of a particular hue appearing within first $k$ peaks and Figure \ref{peaks-cdf-360} shows the cumulative probability distribution of all 360 hues appearing within first $k$ peaks. A particular hue will appear with a probability greater than $99.999\%$ at $k\geq5000$, whereas the probability of all 360 hues appearing at least once in the first 5000 peaks is $99.967\%$. In other words, at 20 peaks per minutes it will take about 250 minutes to capture the highest luminance values of all 360 hues, with a $99.967\%$ success probability.

\subsection{Audio Inference Framework}
\label{inferringsl}
With the above understanding of audio-visualizing light properties, we can now design an inference framework for inferring the source audio from its corresponding luminance-profile. Due to the fundamental differences in the static and random hue modes, we design separate processes for each of them in the inference framework (Figure \ref{attackframework-a}), but both share a common reference library.

\noindent
\textbf{Capturing Luminance-Profile.}
The inference attack starts with the adversary recording the observed luminance-profile ($L_{A_u}'$) of an unknown target song ($A_u$) using the luminance meter. The observed luminance-profile is the time-series of observed luminance values, recorded at a constant sampling rate (10 $Hz$). Therefore, the observed luminance-profile can be represented as $L_{A_u}'=\{\delta_l^{t=1},\delta_l^{t=2},\ldots,\delta_l^{t=m}\}$, where $\delta_l^{t=m}$ is the observed luminance at time instance $m$ since start of the recording. In random hue mode, the adversary additionally records the corresponding hues \linebreak $\delta_h^{t=\{1,2,\ldots,m\}}=\langle \delta_r^{t}:\delta_g^{t}:\delta_b^{t}\rangle$ observed for all $\delta_l^{t=\{1,2,\ldots,m\}}$ in $L_{A_u}'$, for the purpose of normalization.

\noindent
\textbf{Luminance Normalization.}
Once the entire luminance-profile ($L_{A_u}'$) is recorded for a chosen observation duration, the next step for the adversary is to normalize it in order to achieve amplitude invariance during similarity search \cite{batista2011complexity}. In static hue mode, $L_{A_u}'$ is normalized with respect to maximum $\delta_l$ recorded at the point of observation. In random hue mode, the adversary normalizes each observed $\delta_l^{t=\{1,2,\ldots,m\}}$ in $L_{A_u}'$ with respect to the maximum $\delta_l$ recorded at the point of observation for its corresponding hue. The output of this step is a normalized luminance-profile, $\overline{L_{A_u}'}$.

\noindent
\textbf{Creating a Reference Library.} Before matching the normalized luminance-profile ($\overline{L_{A_u}'}$) against a comprehensive reference library of songs, the adversary has to create a reference library of luminance-profiles corresponding to songs in the library of songs. As seen in Figure \ref{B5}, the absolute audio amplitude is directly correlated to the observed luminance. Using this observation, we can create template luminance-profiles by sampling the amplitudes in waveform audio files at 10 $Hz$, and converting them to absolute values. These template luminance-profiles serve as an approximate representation of how a audio-visualizing smart bulb will react.

\noindent
\textbf{Similarity Search.} The final step for the adversary is to match the luminance-profile to songs in reference library, as follows:
\begin{itemize}[leftmargin=*]
\item 
\textit{Dynamic Time Warping.}
We use a classification method based on Dynamic Time Warping (DTW) \cite{muller2007information}, an algorithm for measuring similarity between temporal sequences that are misaligned and vary in time or speed. We compute the DTW distance between the observed luminance-profile and template luminance-profiles in the reference library, selecting the song whose template yields the minimal distance, i.e., we choose the song $i$ such that:

\begin{equation}
A_u = \argmin_{A_i} DTW(\overline{L_{A_u}'}, L_{A_i})
\end{equation}

\item 
\textit{Optimal Subsequence Bijection.}
We also evaluate another classification technique for measuring similarity between temporal sequences, known as the Optimal Subsequence Bijection (OSB) \cite{latecki2007optimal}. OSB is similar to DTW, but allows skipping of elements in the query sequence with a penalty, thus aligning noisy subsequences more efficiently. In our evaluation (Section \ref{sec:audio-visualization-eval}) we present results obtained by using OSB and compare them to those obtained using DTW.
\end{itemize}

\begin{figure}[]
\centering
\includegraphics[width= 0.99\linewidth]{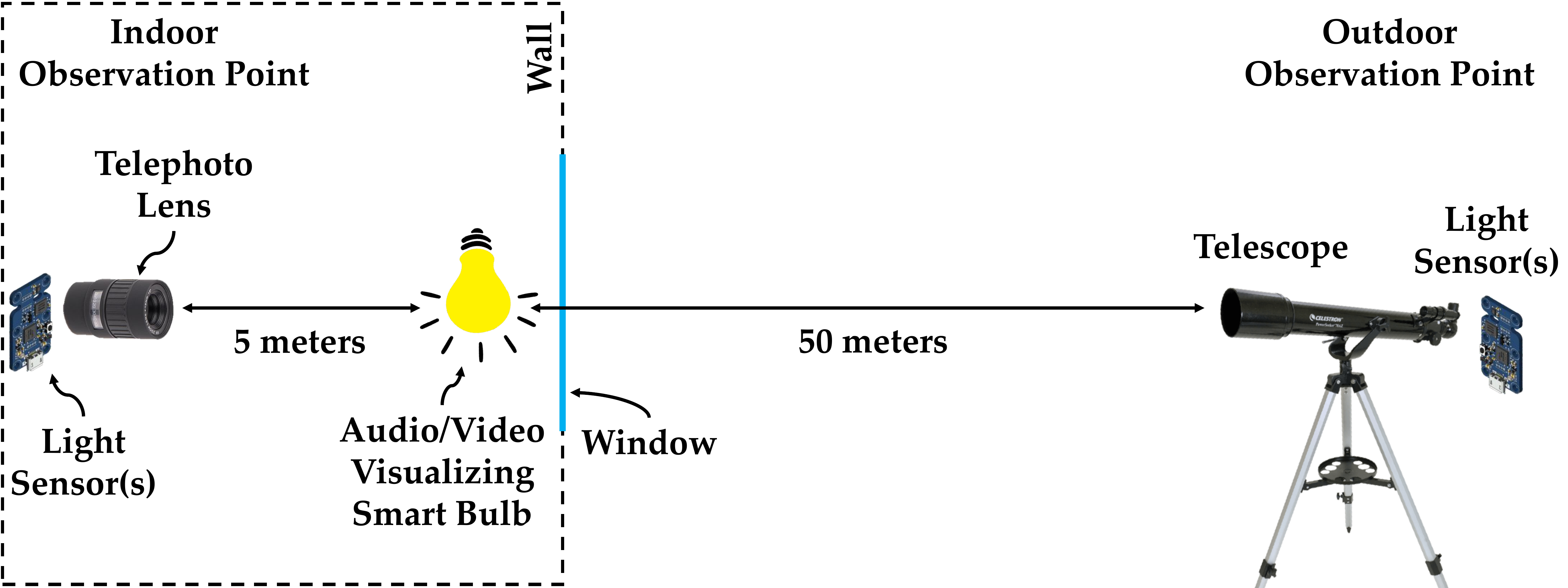}
\caption{Experimental setup for evaluating audio and video inference and data exfiltration attacks.}
\label{expsetup-av}
\end{figure}


\begin{table*}[b]
  \centering
  \scriptsize
  \caption{Mean ranks of 100 test songs in different test settings and observation time window sizes.}
    \begin{tabular}{|r|ccccccccc|cc|} \toprule
          & \multicolumn{9}{c|}{LIFX} & \multicolumn{2}{c|}{Phillips Hue} \\ 
          & \multicolumn{6}{c}{No Obstruction} & 0.8 VT & 0.65 VT & 0.4 VT & \multicolumn{2}{c|}{No Obstruction} \\ 
          & 15 $sec$  & 30 $sec$  & 45 $sec$  & 60 $sec$  & 90 $sec$  & 120 $sec$ & 120 $sec$ & 120 $sec$ & 120 $sec$ & 90 $sec$  & 120 $sec$ \\ \midrule
    Indoor - Static Hue - DTW & 5.47  & 5.35  & 4.24  & 1.90   & 1.18  & 1.01  &  --     &    --   &  -- &    1.23   &  1.09\\
    Indoor - Static Hue - OSB & 5.00     & 3.31  & 2.72  & 1.92  & 1.22  & 1.02  &   --    &   --    &  -- &    1.22   &  1.09\\
    Indoor - Random Hue - DTW & 6.21  & 4.58  & 3.37  & 2.81  & 1.72  & 1.03  &   --    &   --    &  -- &    1.68   &  1.16\\
    Indoor - Random Hue - OSB & 5.52  & 4.79  & 3.57  & 1.91  & 1.21  & 1.03  &     --  &    --   &  -- &    1.24   &  1.17\\
    Outdoor - Static Hue - DTW & 7.57  & 5.48  & 4.13  & 3.06  & 1.78  & 1.20   & 1.63  & 2.95  & 4.47 &    1.79   &  1.35\\
    Outdoor - Static Hue - OSB & 6.69  & 5.45  & 3.21  & 1.99  & 1.50   & 1.19  & 1.65  & 2.42  & 5.10 &    1.62   &  1.56\\
    Outdoor - Random Hue - DTW & 8.67  & 4.35  & 4.03  & 2.11  & 1.53  & 1.15  & 1.84  & 3.64  & 5.85 &    1.77   &  1.27\\
    Outdoor - Random Hue - OSB & 7.80   & 6.35  & 4.62  & 3.06  & 1.93  & 1.21  & 1.81  & 3.67  & 5.79 &  2.01  & 1.25\\ \bottomrule
    \end{tabular}
  \label{time-ranks-mean-a}
\end{table*}

\section{Audio Inference Evaluation}
\label{sec:audio-visualization-eval}
We comprehensively evaluate our audio inference framework (Section \ref{inferringsl}) across different experimental parameter such as observation time duration, distance between adversary and the smart bulb, visible transmittance of a window between the smart bulb and adversary, and using different brands of smart bulbs (LIFX A19 Color Gen3 and Phillips Hue Color Gen3).

\noindent \textbf{Experimental Setup}
We complied a reference library of 400 chart-topping songs, with equal composition of four different genres: country, dance, jazz, and rock. We test two different observation points, one in an indoor setting where the observation point was at 5 meters away from the bulb, another in an outdoor setting where the observation point was at 50 meters away from the bulb (Figure \ref{expsetup-av}). While the outdoor setting represents a more realistic attack scenario for the audio inference attack, the indoor observation point can also be useful 
if the target user is listening using a headphone/earphone (and if the adversary has indoor access).
For the indoor setting, a 20 $mm$ 12x telephoto lens was used to focus light on the sensors. For the outdoor setting, a 80 $mm$ 45-225x telescope was used.

\noindent \textbf{Observation Time and Accuracy}
First we evaluate the success rate of our audio inference framework without any obstruction between the smart bulb and observation point. The window was kept open for the outdoor setting. The LIFX bulb was used in this part of the evaluation, and audio-visualization was performed using the manufacturer's LIFX Android application. For 100 test songs (25 in each of the four genres), we analyzed observation durations varying from 15 to 120 seconds. The time at which the adversary starts observing each test song's audio-visualizing light output was chosen at random. In static hue mode, the hue was chosen at random before the playback of each test song. We measure the accuracy of the framework based on the rank of predicted songs matched against the entire reference library (rank of 1 being the correct prediction). From Table \ref{time-ranks-mean-a}, we can see that the mean rank of predicted songs decreases with a larger observation time window, which is intuitive. 
Inference accuracy in random hue mode was slightly worse than static hue, but conforms to similar improvements for longer observation time windows as static hue. Outdoor inference accuracy was lower than the indoor setting, which is expected due to the lower intensity of light reaching the sensors.
Also, we did not observe a significant advantage of using OSB over DTW. OSB was slightly more accurate than DTW for shorter observation durations, however, their accuracies were very similar for the 120 $sec$ window sizes. 
This can be attributed to the fact that, although OSB has the ability to skip outliers in the test sequence, we have very few outliers in our test sequences to begin with. This is because both audio-visualization and sensor sampling follow approximately constant sampling rates.

\begin{figure}[b]
\centering
\includegraphics[width=0.99\linewidth]{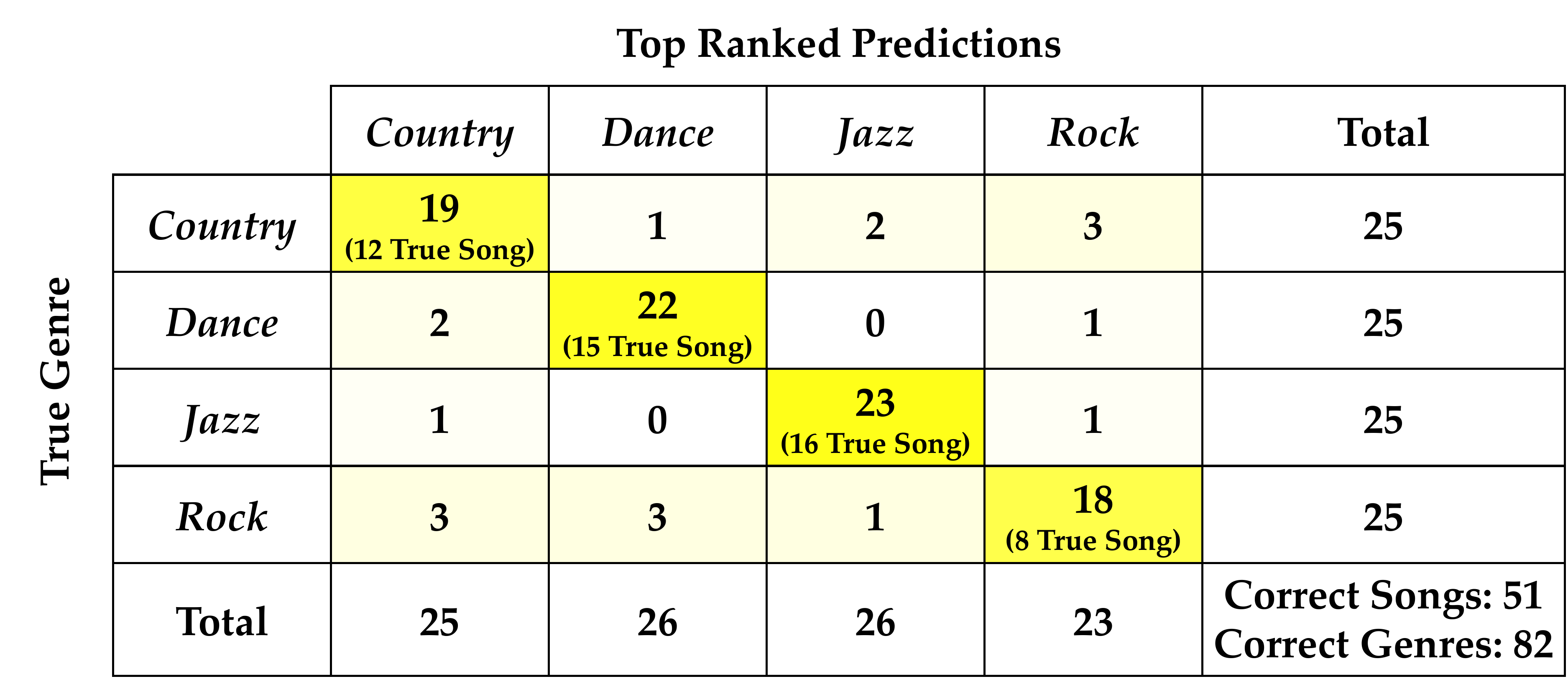}
\caption{Genre confusion matrix of 100 test songs.}
\label{genres}
\end{figure}

\noindent \textbf{Music Genres}
Certain genres are slower paced (such as jazz) than others (such as dance). Some genres have high frequency periodic beats (such as dance) while others are dominated with flat vocals (such as country). These characters are also reflected on the audio-visualizing smart lights, and results in a higher confusion within the same genre. Figure \ref{genres} shows the genre confusion matrix of 100 songs tested under the following settings: 60 $sec$ observation duration, indoor, static hue, OSB (from results in Table \ref{time-ranks-mean-a}). We see that the highest confusion is more likely to happen within the genre of the test song. In this example, 51 songs were correctly predicted in the top rank, while genres of 82 songs were correct in the same prediction. This trend was observed across all different test scenarios. This implies that even if the adversary is unable accurately match the audio-visualized data to its corresponding song, he/she can still infer the user's media or genre preference.

\noindent \textbf{Visible Transmittance of Window}
We next evaluate how our audio inference framework performs when there is partial obstruction, such as a tinted window, between the smart bulb and the adversary. We used three window glasses of varying Visible Transmittance (VT) -- approximately 0.8, 0.65 and 0.4 -- in the outdoor setting and evaluated the prediction accuracy. 
We repeated the outdoor test with 100 songs using an 120 $sec$ observation time window size. In Table \ref{time-ranks-mean-a}, we see that the mean ranks of the predicted song drops as the visible transmittance of the window decreases. For example, the mean rank of predicted song was 1.20 for inference using an open window (test settings: outdoor, static hue, DTW), which declined to 1.63 using a 0.8 VT window, 2.95 using a 0.65 VT window, and 4.47 using a 0.4 VT window. This is expected because with lower VT of the window, the intensity of light reaching the sensors is diminished. Thus, highly tinted windows can act as an effective protection measure against such an inference attack.

\noindent \textbf{Generalizability}
We also evaluate the generalizability of our results using other smart bulbs. We chose the Phillips Hue bulb for this evaluation because of it's popularity and it is currently one of the few smart bulbs that support audio-visualization. However, due to the lack of a native audio-visualization feature in the Phillips Hue mobile application, audio-visualization was done using hueDynamic, a third-party application. Table \ref{time-ranks-mean-a} shows the mean ranks of 100 test songs predicted after observing the audio-visualization of the Phillips Hue bulb for 90 and 120 seconds. The mean ranks in most settings, indoor or outdoor, were slightly higher than that of using the LIFX bulb. On an average across all different test settings, the mean rank of the predicted song increased by 0.06 and 0.13 for 90 $sec$ and 120 $sec$ observation time windows, respectively. While these results validate the generalizability of our inference attack, we investigated why the accuracy was consistently lower than when using the LIFX bulb. One of the primary contributing factors may be the difference in the maximum light output between the two bulbs. The LIFX A19 Gen3 bulb is rated at a maximum of 1100 $lm$ while the Phillips Hue Gen3 is rated at a maximum of 800 $lm$. As a result, the intensity of light reaching the sensors is lower, and aligning with our previous observations, this reduces the accuracy of our inference framework.


\section{Video Inference Threat}
\label{sec:video-visualization}
\textit{When video-visualization is turned on in the mobile app, the smart bulb reacts to the colors present in the input video stream by changing its output light color to the average RGB composition of the current frame in the video}. Unlike audio-visualization which is better understood with the HSB color model, video-visualization is better represented using the RGB model. This is because in audio-visualization the saturation level never changes, whereas saturation (along with hue and brightness) in video-visualization is entirely dependent on the color composition of the current video frame. As all three variables are dynamic, using the RGB model simplifies the matching process with responses on a RGB sensor, as described below.

\subsection{Effect of Video on Bulb Color}
\begin{figure}[]
\centering
\begin{subfigure}{0.49\linewidth}
\centering
\includegraphics[width=\textwidth]{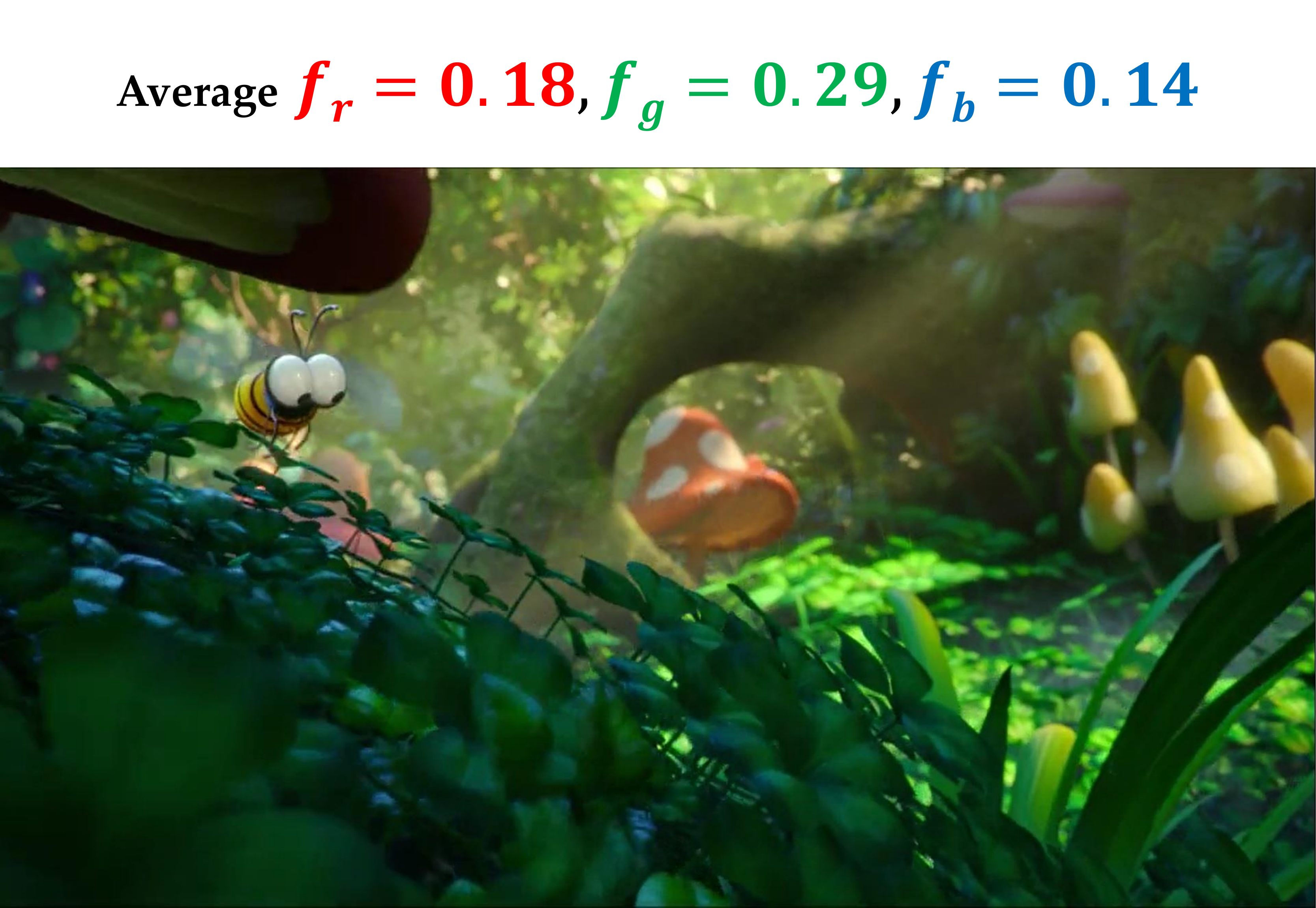}
\caption{}
\label{video-sample}
\end{subfigure}
\hfill
\begin{subfigure}{0.49\linewidth}
\centering
\includegraphics[width=\textwidth]{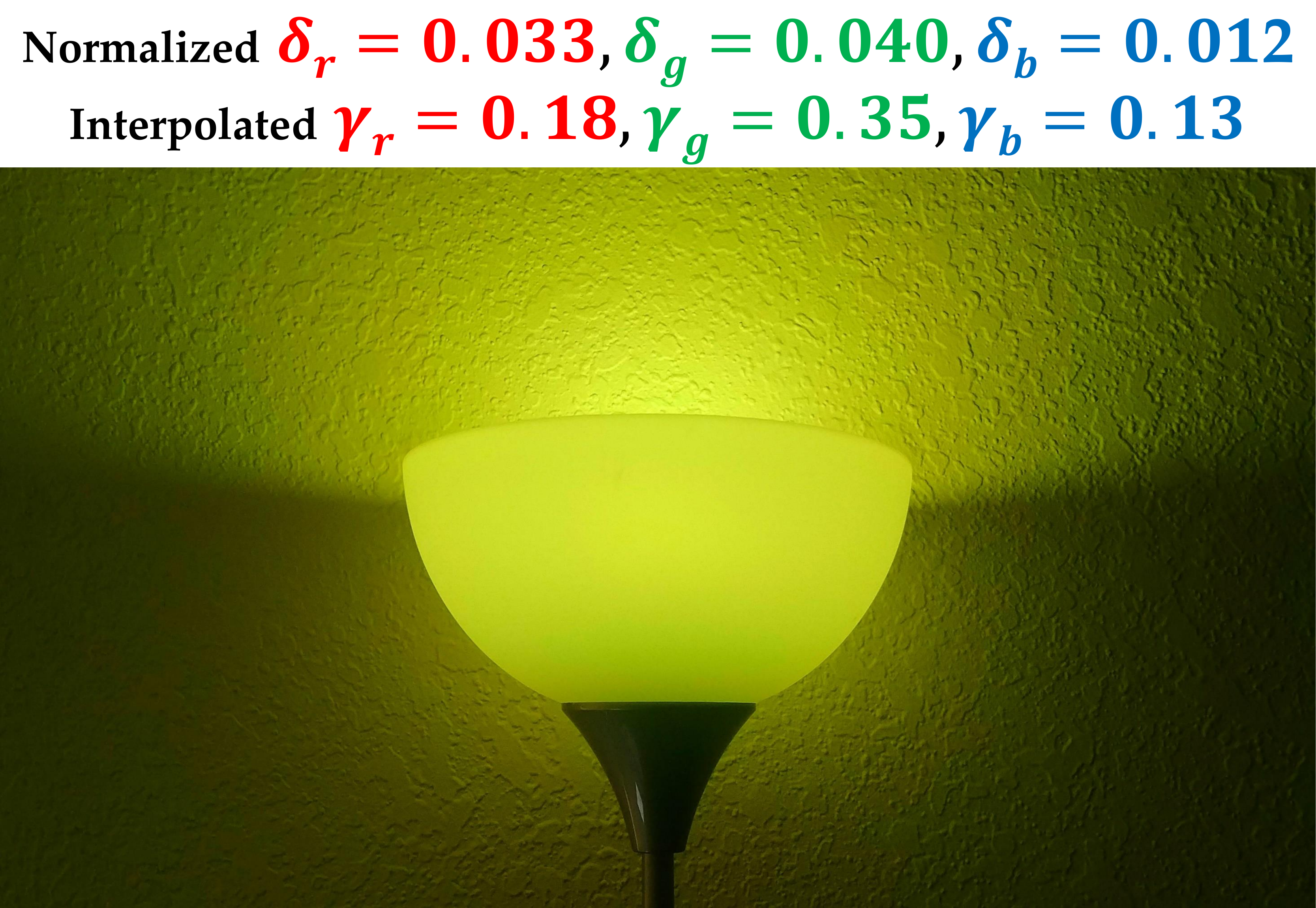}
\caption{}
\label{video-sample-light}
\end{subfigure}
\caption{(a) -- A frame from the movie \textit{Smurfs: The Lost Village (2017)}; (b) -- Corresponding light output observed on a video-visualizing LIFX bulb, $\langle \delta_r,\delta_g,\delta_b\rangle$.}
\label{video-sample-and-light}
\end{figure}

To precisely study the video-visualization properties, we created another exploratory setup (Appendix B). We played a few sample video files multiple times and observed the corresponding time-series of bulb's RGB color measured using the Vernier GDX-LC RGB sensor. The GDX-LC's RGB sensor captures responses in the 615 $nm$, 525 $nm$, and 465 $nm$ peak wavelength ranges, approximately representing the observed intensities of red, green and blue colored lights, respectively. Our observations brought us to two similar conclusions as in section \ref{audio-effect}. First, the observed RGB color of the bulb $\langle \delta_r,\delta_g,\delta_b\rangle$ has some correlation with the average RGB color in the current frame $\langle f_r,f_g,f_b\rangle$ of the video stream (Figure \ref{video-sample-and-light}), which is expected. As a result, the RGB ``color-profile'' of a video (the time-series of $\langle \delta_r,\delta_g,\delta_b\rangle$ when the video is playing) observed on a video-visualizing bulb is unique to the video, and the probability that a completely different video also has the same color-profile is small (due to the high number of colors possible per pixel, in each frame of both videos). 
Second, similar to luminance-profiles in audio-visualization, color-profiles also suffers minor distortions across multiple recordings. If the color-profile of a video $V_1$ captured during two different playbacks is denoted by $C_{V_1}'$ and $C_{V_1}''$, $C_{V_1}'$ and $C_{V_1}''$ will be similar but not identical. These minor distortions can be attributed to varying network latencies, packet loss due to network congestion and  varying RGB calculation time. However, similarity between $C_{V_1}'$ and $C_{V_1}''$ is generally good, which can be utilized for designing dictionary-based elastic time-series matching attacks. 

\subsection{Video Matching Using Color-Profile}
\label{video-interpolation-logic}
We already detailed some characteristics of RGB smart bulbs in random hue audio-visualization (Section \ref{audio-matching}). However, identifying hues based on observed response ratios $\langle \delta_r:\delta_g:\delta_b\rangle$ is relatively simpler than accurately identifying a $\langle r,g,b\rangle$ color from the observed responses. This is because all three variables (hue, saturation and brightness) of the HSB model are dynamic in video-visualization. Moreover, the non-ideal characteristics of the light output, and/or RGB sensor, makes it challenging to derive an accurate relation between $\langle f_r,f_g,f_b\rangle$ and $\langle \delta_r,\delta_g,\delta_b\rangle$. The two most significant problems in inferring $\langle f_r,f_g,f_b\rangle$ from $\langle \delta_r,\delta_g,\delta_b\rangle$ are: (i) the primary colors have different peak responses (Figure \ref{rgb-cd-cycle}), and (ii) the responses are non-linear with respect to brightness (Figure \ref{white-rgb}). We address both of the above problems by first normalizing all observed $\delta_r$, $\delta_g$ and $\delta_b$ with respect to the highest observable $\delta_r$, $\delta_g$ or $\delta_b$. In case of both LIFX and Phillips Hue bulbs, $\delta_b$ has the highest observable response at 100\% brightness (Figure \ref{white-rgb}). This step also addresses other variable factors such distance from the light source, and optical amplifications before the sensor. After normalization, the $\delta_r$, $\delta_g$ and $\delta_b$ are interpolated based on non-linearities learned for each of the primary colors (Appendix A). The interpolated response $\langle \gamma_r,\gamma_g,\gamma_b\rangle$ is much closer to the true composition $\langle f_r,f_g,f_b\rangle$ of the current frame (Figure \ref{video-sample-and-light}). Even after normalization and interpolation, $\langle f_r,f_g,f_b\rangle$ and $\langle \gamma_r,\gamma_g,\gamma_b\rangle$ will not be an exact match for most colors, because the observable color responses are not to ideal RGB proportions (Figure \ref{rgb-cd-cycle}). However, $\langle f_r,f_g,f_b\rangle$ and $\langle \gamma_r,\gamma_g,\gamma_b\rangle$ are similar enough that we can create a reference library of color-profiles for a diverse set of videos, and use it to match the observed color-profile of a target video.

\subsection{Video Inference Framework}
\label{inferringvl}
With the above understanding of video-visualizing light properties, we can now design an video inference framework.

\noindent
\textbf{Capturing Color-Profile.}
The inference attack starts with the adversary recording the observed color-profile ($C_{V_u}'$) of an unknown target video ($V_u$) using the color sensor. The observed color-profile is the time-series of observed RGB values, recorded at a constant sampling rate (1 $Hz$). Therefore, the observed color-profile can be represented as $C_{V_u}'=\{\langle \delta_r,\delta_g,\delta_b\rangle^{t=1},\langle \delta_r,\delta_g,\delta_b\rangle^{t=2},\ldots,$ $\langle \delta_r,\delta_g,\delta_b\rangle^{t=m}\}$, where $\langle \delta_r,\delta_g,\delta_b\rangle^{t=m}$ is the observed luminance at time instance $m$ since start of the recording.

\noindent
\textbf{Color Normalization and Interpolation.}
Once the entire color-profile ($C_{V_u}'$) is recorded for a chosen observation duration, the adversary next normalizes and interpolates (as described in Section \ref{video-interpolation-logic}) it to create the corresponding normalized and interpolated color-profile ($\overline{C_{V_u}'}$).

\noindent
\textbf{Creating a Reference Library.} 
The adversary next creates a library of template color-profiles corresponding to video files in a reference library. 
The template color-profiles can be created by sampling the RGB composition in video files at 1 $Hz$. These template color-profiles serve as an approximate representation of how a video-visualizing smart bulb will react (Figure \ref{video-sample-and-light}).

\noindent
\textbf{Similarity Search.} The final step for the adversary is to match the color-profile ($\overline{C_{V_u}'}$) against the template color-profiles corresponding to the reference library of videos using a time-series similarity computing technique such as \textit{Multidimensional Dynamic Time Warping (MDTW)} \cite{wollmer2009multidimensional} . MDTW is a generalization of DTW for measuring similarity between temporal sequences, in two or more dimensions. We compute the 3-DTW distance between the observed color-profile and template color-profiles in the reference library, selecting the video whose template yields the minimal distance. More formally, we choose the video $i$ such that:

\begin{equation}
V_u = \argmin_{V_i} MDTW(\overline{C_{V_u}'}, C_{V_i})
\end{equation}

\section{Video Inference Evaluation}
\label{sec:video-visualization-eval}
We comprehensively evaluate the performance of our video inference framework (Section \ref{inferringvl}) for a variety of parameters. However, as video-visualization shares some of the same fundamental properties as audio-visualization, we expect to see similar trends in inference accuracy for factors such as window's visible transmittance and use of different bulbs. Therefore, here we investigate other factors of interest such as the size of the reference library.

\noindent \textbf{Experimental Setup}
We complied a reference library of 500 full-length movies released on blu-ray in the last 10 years. We test from the same two observation points as in the audio inference evaluation, indoor and outdoor (Figure \ref{expsetup-av}). Also, the same telephoto lens and telescope were used in the indoor and outdoor settings, respectively. A LIFX A19 bulb was used for video visualization, and the GDX-LC RGB sensor was used to record the bulb's color output.

\begin{figure}[]
\centering
\includegraphics[width=0.89\linewidth]{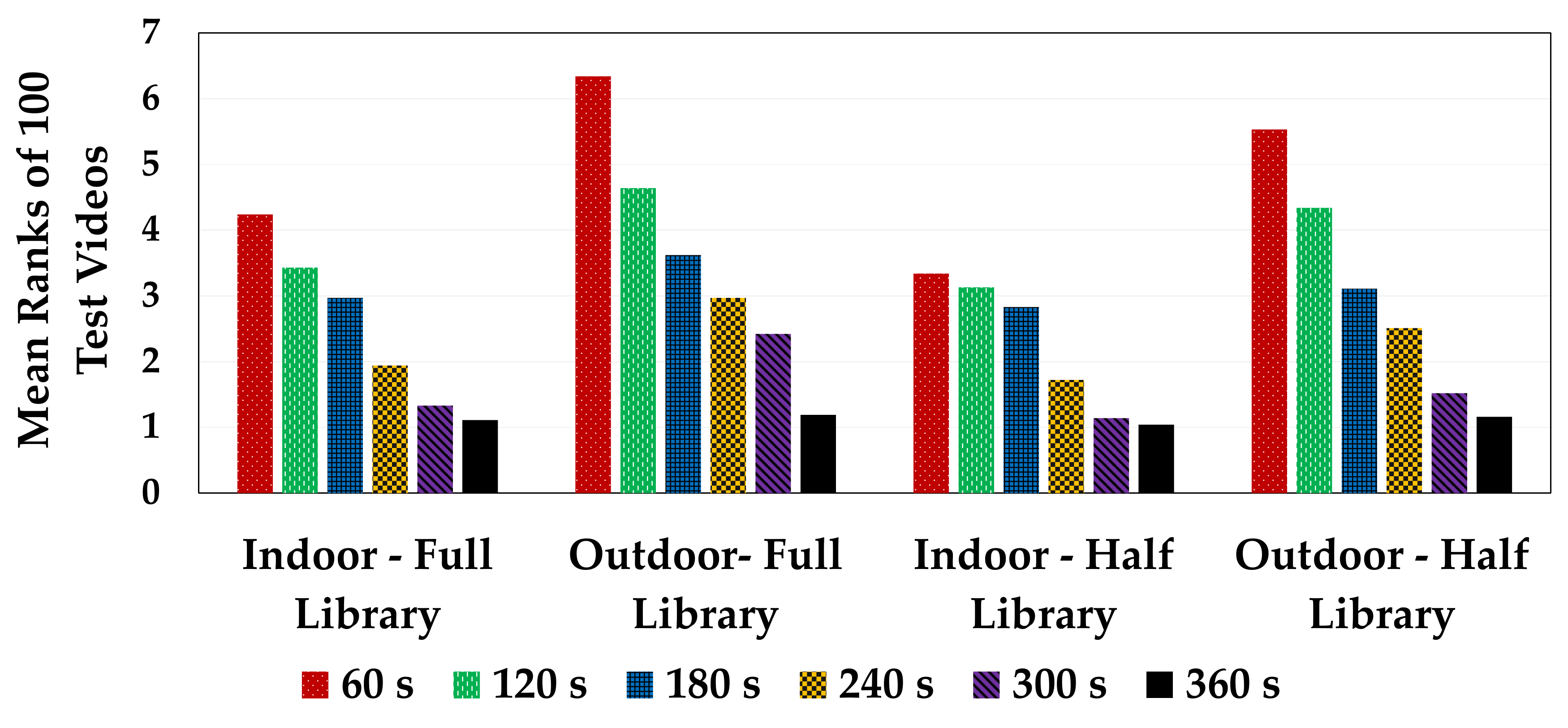}
\caption{Mean ranks of 100 test videos in different settings.}
\label{time-ranks-v}
\end{figure}

\noindent \textbf{Observation Time, Distance and Accuracy}
Similar to audio inference, we calculate the video inference accuracy based on the mean rank of 100 test videos. For each of the 100 test videos, we analyzed observation durations varying from 60 to 360 seconds. The time at which the adversary starts observing each test video's video-visualizing light output was chosen at random. In Figure \ref{time-ranks-v} we see that the mean rank of predicted test videos decreases with a larger observation time window. For example, the mean rank was 6.34 for inference using a 60 $sec$ observation time window (outdoor observation), which improved to 1.19 using a 360 $sec$ window. Also, indoor accuracy was more accurate with mean rank of 4.24 for a 60 $sec$ window to 1.11 using a 360 $sec$ window.

\noindent \textbf{Size of the Reference Library}
Larger dictionary increases the potential of a mismatch with another video having a segment with similar color-profile as the observed segment. We validate this using only half of the reference library for prediction (250 movies). In Figure \ref{time-ranks-v} we see that the mean rank of predicted test videos decreases with a smaller reference library. For example, the mean rank was 6.34 for inference using 60 $sec$ observation window (outdoor observation) and the full library, which improved to 5.53 using a 60 $sec$ window and half the library. 

\section{Covert Data Exfiltration Threat}
\label{sec:infrared-attack}
\begin{figure}[]
\centering
\includegraphics[width=\linewidth]{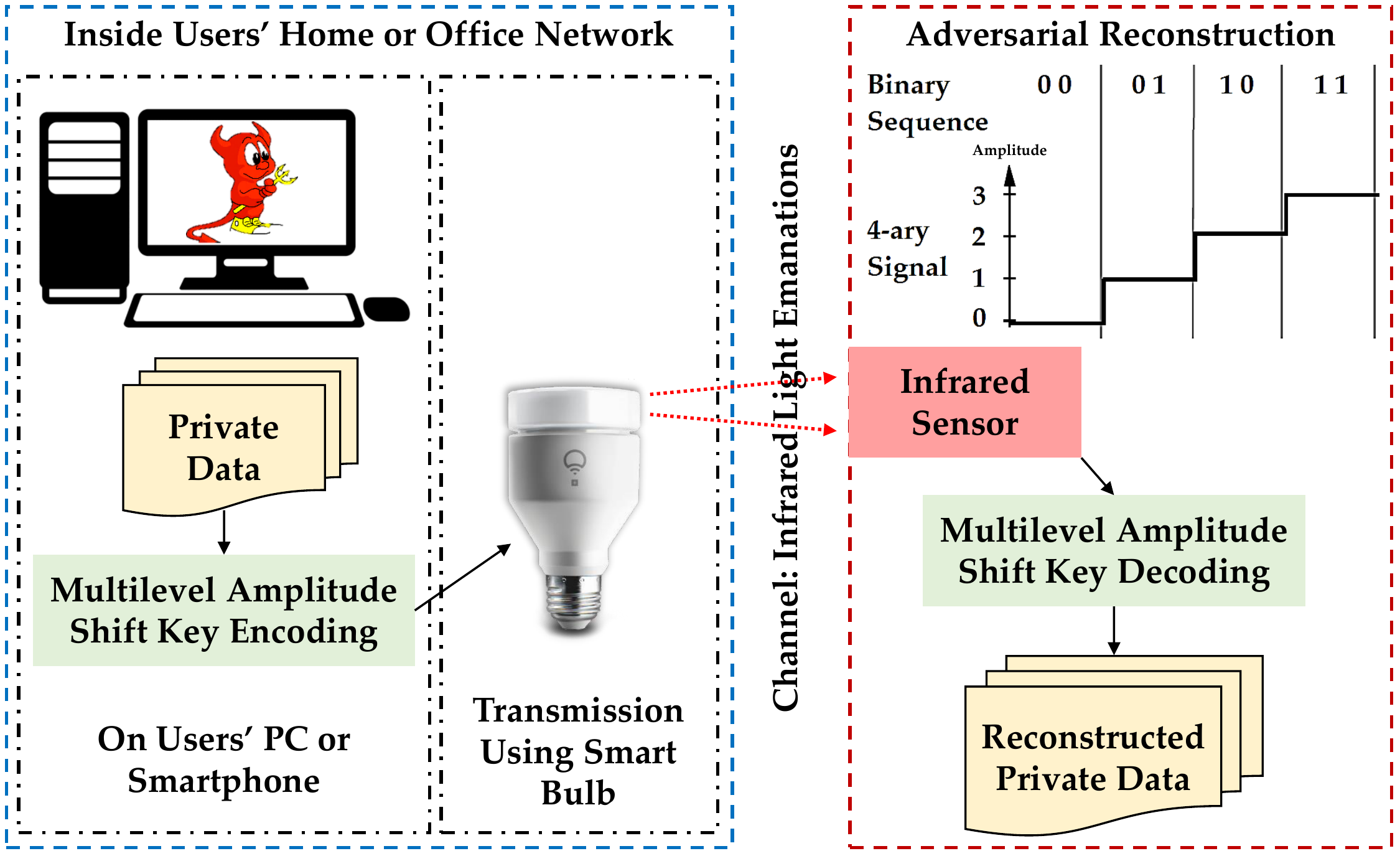}
\caption{Smart light-based data exfiltration framework.}
\label{attackframework-i}
\end{figure}

Next we present another smart light based attack framework, using which an adversary can actively and covertly exfiltrate private data from within a smart light user's personal device or network. This attack is possible on hub-less smart lights (e.g., LIFX) due to the lack of permission controls for controlling the light within the local network. This attack may also be possible using smart lights that connect via a hub, only if the hub does not have permission controls. However, the Phillips Hue ecosystem uses a hub with permission controls, and thus cannot be used for this attack unless the malicious software agent obtains access permissions by masquerading as an useful application (a Trojan). However, Phillips Hue currently does not offer any infrared-enabled smart lighting products, so we use the LIFX ecosystem to design and evaluate the proposed data exfiltration framework (Figure \ref{attackframework-i}). 

\subsection{The Covert Channel}
The adversary can potentially use either, or both, the visible and infrared spectrum of a smart bulb to create an one-way line-of-sight data exfiltration gateway, using transmission techniques such as amplitude and/or wavelength\footnote{As the LIFX+ series supports only 950 $nm$ infrared light, only amplitude shift keying is possible with the infrared spectrum.} shift keying \cite{xiong2005amplitude,monteiro2014design}. However, uninitiated changes in the visible light amplitude or color is more likely to get noticed, thus limiting the effectiveness of such a malicious gateway/channel. As human eyes are not sensitive to the infrared spectrum, it can be used to create a covert channel which can remain undetected for longer durations.

Depending on the amount of private data to be covertly exfiltrated, channel bandwidth can be a deciding factor in the success of such an attack. As a higher bandwidth channel will take less time to transmit the data, compared to a lower bandwidth channel, the likelihood of a detection and/or disruption is lessened with higher bandwidth.
In our empirical tests we observed that LIFX+ bulbs can achieve a maximum infrared output (power level 65535) from an off state (power level 0) in approximately 0.45 seconds, and takes less than 0.2 seconds to completely turn off from a maximum output state. As these are the maximum switching response times, we can use the higher of the two as our clock period. After adding some padding we round up the clock period to 0.5 seconds ($ClockRate=2$ $Hz$), which results in the maximum channel bandwidth of 32 bits per second if a multilevel ($M$-ary) amplitude shift keying (ASK) \cite{avlonitis2006multilevel} encoding technique is employed with $M=65536$. Channel bandwidth in a $M$-ary ASK is $(\log_{2} M)\times{ClockRate}$.
 However, distinguishing between each of the 65536 power levels (amplitudes) may be difficult, especially, if the distance of the receiver from the smart bulb is large, primarily due to signal attenuation and channel noise. As a result, the error rate at the receiver is to some extent proportional to $M$ and the distance. 

\subsection{Private Data Encoding and Transmission}
Once the value of $M$ is decided by the adversary (based on the distance and acceptable level of reconstruction quality), the next step is to encode private data of interest (in binary form) using $M$-ary ASK. This task is undertaken by a malicious software agent installed on the target user's device or network, as outlined earlier in Section \ref{sec:adversarymodel}. The encoded data is then transmitted in blocks by controlling the infrared power level of the smart bulb connected to the same device or network. For example in 4-ary ASK, a `00' data block will be transmitted by setting the infrared power level of the bulb at 0 (off), a `01' data block will set the power level  at 21845 ($65536\div(M-1)$), a `10' data block will set the power level at 43690, and a `11' data block will set the infrared power level at 65535 (maximum). Before data transmission, a pre-decided start symbol \cite{porter1987system} is used to activate the receiver, synchronize the clock, and set maximum amplitude for normalization. After data transmission is complete, an end symbol is used to signal termination. 

\subsection{Adversarial Reconstruction of Private Data}
On the adversary's side, he/she observes the target user's smart bulb using an infrared sensor. As the LIFX+ bulbs supports only 950 $nm$ infrared light, the adversary requires compatible infrared sensing hardware. We used a TSOP48 \cite{TSOP48} infrared sensor and recorded data using an ATtiny85-based Arduino board. Once a start symbol is received by the infrared sensor, the adversary starts recording the observed infrared amplitudes representing the M-ary ASK encoded data, until an end symbol is received. Then the adversary normalizes the recorded data based on the maximum amplitude, and decodes it to reconstruct the private data in binary format. Depending on the amount of signal attenuation and channel noise, the reconstructed data may not be identical to the original data. 

\section{Data Exfiltration Evaluation}
\label{sec:infrared-attack-eval}
Next, we present performance evaluation results of our data exfiltration framework outlined earlier.

\begin{figure}[]
\centering
\begin{subfigure}{0.19\linewidth}
\centering
\includegraphics[width=\textwidth]{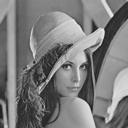}
\caption{Original}
\label{lena-original}
\end{subfigure}
\begin{subfigure}{0.19\linewidth}
\centering
\includegraphics[width=\textwidth]{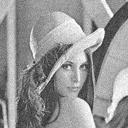}
\caption{5 $m$}
\label{lena-5m}
\end{subfigure}
\begin{subfigure}{0.19\linewidth}
\centering
\includegraphics[width=\textwidth]{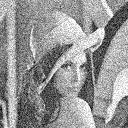}
\caption{15 $m$}
\label{lena-15m}
\end{subfigure}
\begin{subfigure}{0.19\linewidth}
\centering
\includegraphics[width=\textwidth]{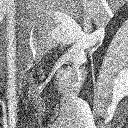}
\caption{30 $m$}
\label{lena-30m}
\end{subfigure}
\begin{subfigure}{0.19\linewidth}
\centering
\includegraphics[width=\textwidth]{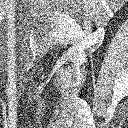}
\caption{50 $m$}
\label{lena-50m}
\end{subfigure}
\caption{Reconstructed image of Lena at different distances, using 128-ary ASK.}
\label{lena}
\end{figure}

\noindent \textbf{Experimental Setup}
Standardized datasets -  first 10 sets of Harvard sentences \cite{rothauser1969ieee} and a 128$\times$128 image of Lena (Figure \ref{lena-original}) - were used to evaluate the Bit Error Rate (BER) in the reconstructed data. The adversarial observation distance from the smart bulb was increased in steps (5, 15, 30, and 50 meters), and BER was calculated for different values of $M$ (64, 128, 256, 512, 1024, and 2048) at each of these distances. A LIFX+ A19 Gen3 bulb was used for transmission and a 80 $mm$ 45-255x telescope was used to focus light on the infrared sensor (TSOP48). A python script, acting as the malicious software agent, was used to encode and transmit the test data. It was executed on a Windows PC connected to the same Wi-Fi network as the bulb.

\begin{figure}[]
\centering
\includegraphics[width=0.8\linewidth]{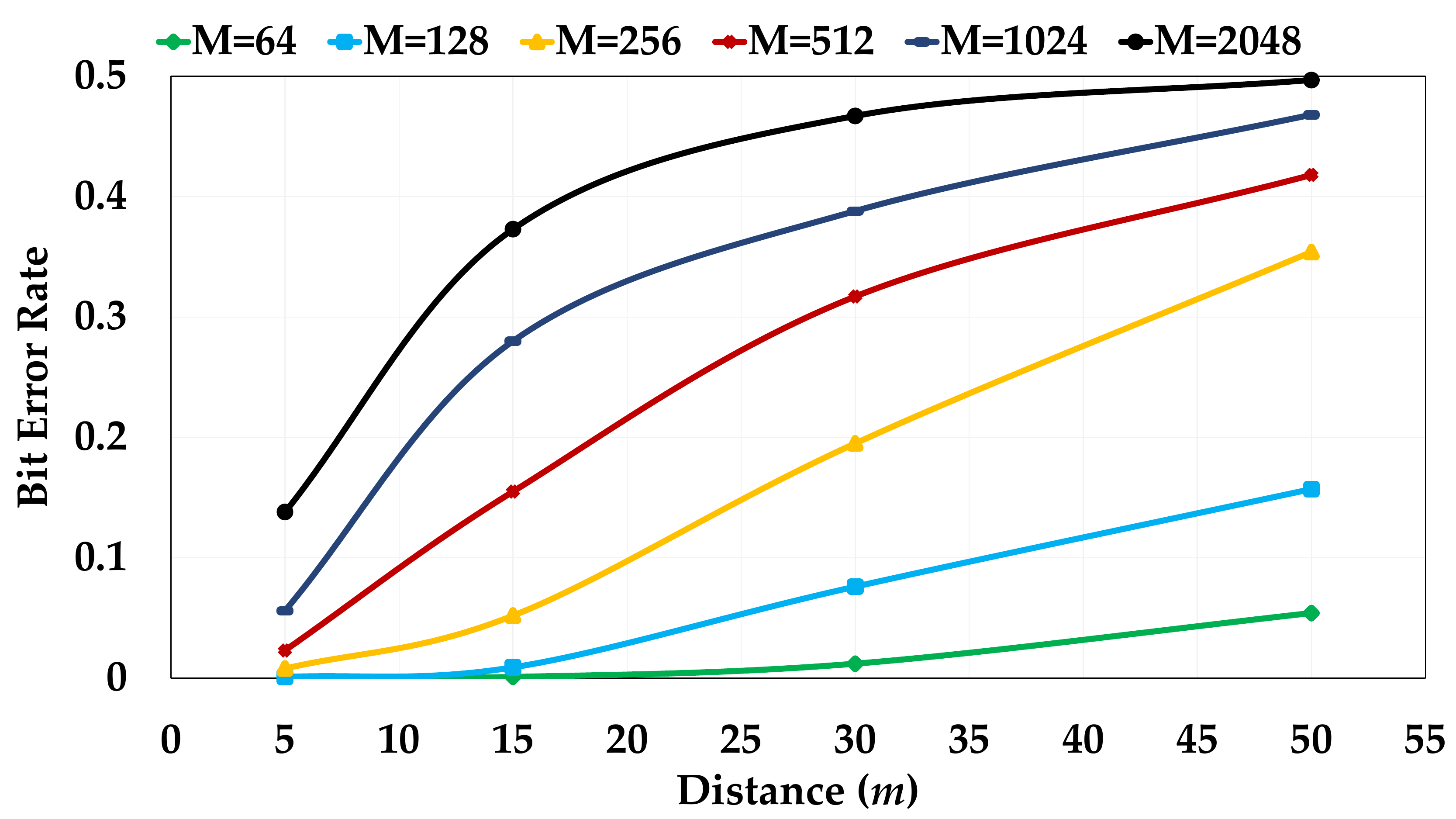}
\caption{Mean BER in exfiltration of both text and image.}
\label{ir-ber}
\end{figure}

\noindent \textbf{Distance and Error Rate}
The infrared signal strength reduces as distance from the bulb increases. As a result, the boundaries between the $M$ amplitude levels present in a $M$-ary ASK signal is also diminished, leading to higher confusion between neighboring amplitude levels and thus errors in the reconstructed data. This phenomenon was evident in our evaluation results (Figure \ref{ir-ber}), especially for higher values of $M$. For example, with $M=2048$ the BER increased from 0.138 to 0.496 when the distance was increased from 5 $m$ to 50 $m$. Note that when the signal is very week or corrupted, the BER converges to 50\%, provided that the binary data source used approximately follows a fair Bernoulli distribution. In Figure \ref{lena} we see how the image of Lena is degraded with higher BER for longer distances.

\begin{figure}[b]
\centering
\includegraphics[width=0.99\linewidth]{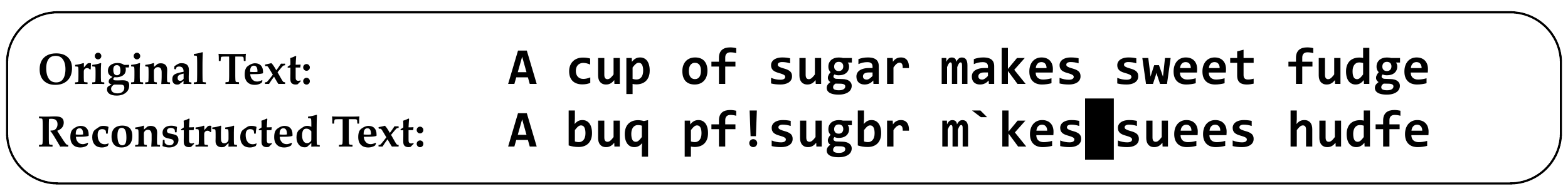}
\caption{Text reconstruction using 256-ary ASK at 15 $m$.}
\label{ir-text-example}
\end{figure}

\noindent \textbf{$M$ and Error Rate}
The boundaries between the $M$ amplitude levels present in a $M$-ary ASK signal is also diminished when the value of $M$ is increased, leading to higher confusion between neighboring amplitude levels. In Figure \ref{ir-ber}, we observe that the BER consistently increases for higher values of $M$. However, even with high BER the reconstructed data can still be useful for an adversary. For example, text inference from reconstructed data is easier than many other encoding schemes, because of the higher likelihood of confusion between neighboring amplitude levels. Figure \ref{ir-text-example} shows the example of a reconstructed Harvard sentence using 256-ary ASK. We see that the incorrectly reconstructed letters are neighboring to the original letters on the ASCII chart. For example, the blank-space character (ASCII value 32) was confused with `!' (ASCII value 33) and unit-separator (ASCII value 31). An adversary can perform additional semantic analysis on the reconstructed text to improve its correctness and legibility.

\section{Related Work}
\label{sec:related}

\noindent
\textbf{Private Information Leakage through Optical Side-channels.}
Various forms of private information leakage threats that employ an optical side-channel have been demonstrated in the literature. 
For instance, Xu et al. \cite{xu2014watching} were able to infer the video being watched using the changing light characteristics observable through the target's window.
Similarly, Schwittmann et al. \cite{schwittmann2016video} were able to infer a video played on a TV, by exploiting a smartphone's ambient light sensor.
An extended work by Schwittmann et al. \cite{schwittmann2016identifying} employed the ambient light sensor on a smartwatch to accomplish the same attack. 
Alternatively, Backes et al. \cite{backes2009tempest,backes2008compromising} demonstrated how reflections from objects such as, users' eyes, teapots and walls can be exploited for spying on printed or digitally displayed data. 

The first inference threat that we demonstrated in this paper employed an optical side-channel to infer the source audio. While there are multiple previous research efforts \cite{wei2015acoustic,roy2016listening,davis2014visual,savage2015visualizing,zhang2015accelword,michalevsky2014gyrophone,han2017pitchln} on inferring audio from different types of information side-channels, our use of optical emanations from multimedia-visualizing smart lights  as an information side-channel for this task is unique and novel. 
Our second inference threat employed an optical side-channel to infer the video being watched by the target user is similar to works by Xu et al. \cite{xu2014watching} and Schwittmann et al. \cite{schwittmann2016video}. However, our video inference framework which targets optical emanations from a multimedia-visualizing smart light is significantly different from Xu et al.'s and Schwittmann et al.'s inference frameworks that target light emanations from a TV screen. As smart bulbs are much brighter than TV screens, they make a desirable target for distant outdoor attackers. However, the distinguishing features in a smart bulb's output is often limited to only one color (in the RGB space) at a time as outlined in Section \ref{sec:background}. This makes it more challenging to accurately infer the source video from a smart light's output, compared to a more straightforward analysis of the various simultaneous features present in a TV screen's output.

\noindent
\textbf{Information Exfiltration through Optical Covert-channels.} 
Loughry et al. \cite{loughry2002information} were one of the first to call attention to information exfiltration attacks on air-gapped systems by employing visible-light LED indicators to transmit bits of information. Zhou et al. \cite{zhou2018irexf} demonstrated a similar attack, but using a malicious infrared signaling hardware installed on the air-gapped system. Guri et al. \cite{guri2016optical} leveraged on the limitations of human visual perception in order to covertly exfiltrate data through a standard computer LCD display, even in the presence of a user. Guri et al. \cite{guri2017air} extended their previous work to exploit cameras equipped with infrared lights to both exfiltrate and infiltrate data from/to an air-gapped network. Shamir \cite{uk2014light} demonstrated how a scanner can be used to infiltrate data/malware in to an air-gapped network. 


A common limitation of the above data exfiltration attacks is that the covert-channel's bandwidth is restricted due to the binary nature of the signaling light, and slow switching response time between individual bits. For example, Zhou et al. \cite{zhou2018irexf} were able to achieve a raw throughput of 2.62 bits per second, even after installing a custom infrared signaling hardware on the air-gapped system. Guri et al.'s \cite{guri2016optical} method of encoding data in the form of QR-codes can potentially achieve relatively higher throughput, but their reconstruction was successful only from a short distance of up to 8 meters. In contrast, infrared-enabled smart lights can act as a superior data exfiltration gateway because - (a) they have fine-grained control of brightness/intensity, which can be used to design communication protocols that achieve higher throughput, (b) they are brighter than LED indicators found on computers and routers, increasing the possibility of data reconstruction from a longer distance, and (c) the adversary does not have to surreptitiously place any additional malicious hardware in the target area (i.e., in addition to the smart light already installed by the user).

\section{Discussions}
\label{sec:discussions}

\noindent \textbf{Limitations and Improvements}: There are several factors that can limit or improve the attacks presented in this paper. Factors such ambient light (noise), reflections off a color painted wall, no line-of-sight access to target user's window, etc. can limit the success of these attacks. On the other hand, an adversary can improve the attacks by using high quality sensors (such as high speed digital cameras) and optical lenses (such as wide aperture telescopes). Our goal in this work was to demonstrate the feasibility of these attacks using low-cost off-the-shelf sensors and lenses (for example, the TSOP48 sensor costs less than \$10).

\noindent \textbf{Countermeasures}:
Concerned smart light users can take several measures against the proposed threats. We already saw that windows with low visible transmittance causes the attacks to perform poorly. Therefore, a simple mitigation would be to cover the windows with opaque curtains and block light leakage to the outside. For the inference attacks, the maximum brightness of the bulbs can be reduced, so that the  light leakage is also reduced. To prevent the exfiltration attack, strong network rules can be enforced such that computers and smartphones cannot control smart bulbs over an IP network. However, such rules may harm the utility of smart bulbs.

%
%
%
%
%


\section{Conclusions}
\label{sec:conclusions}

In this paper, we designed and evaluated two attack frameworks to infer users' audio and video playback by a systematic observation and analysis of the multimedia-visualization functionality of modern smart lights. We also designed and evaluated a covert and high bandwidth data exfiltration attack by taking advantage of the infrared capabilities of these smart lighting systems. We conducted a comprehensive evaluation of these attacks in various real-life settings which confirmed the feasibility of proposed privacy threats. These threats also affirm the need for new protection mechanisms such as better access control in light management applications.

\bibliographystyle{abbrv}
\bibliography{lightears-bibliography} 


\section*{Appendix A - Polynomial Interpolations for Non-Linear Color Responses} 
The observed RGB responses ($\delta_r$, $\delta_g$ and $\delta_b$) have non-linear relationships with brightness (Figure \ref{white-rgb} and Table \ref{white-rgb-table}). Therefore, we employ Lagrange polynomials to learn the non-linearities. We use five observed $\langle \delta_r,\delta_g,\delta_b\rangle$ points for $RGB\langle 1:1:1\rangle$  in Table \ref{white-rgb-table} to calculate Equation \ref{lagrange-rgb}, using Lagrange basis polynomials (Equation \ref{lagrange-poly-rgb}). The functions in Equation \ref{lagrange-rgb} are then used in the attack phase (with different distance and amplification factors) to accurately interpolate all observed $\delta_r$, $\delta_g$ and $\delta_b$ values in a color-profile. A similar set of functions were also learned and applied for attacks on the Phillips Hue bulb. $\gamma_{r}$, $\gamma_{g}$ and $\gamma_{b}$ remain approximately constant across different amalgamations of primary colors, for both LIFX and Phillips Hue.

{
\begin{equation}
\begin{aligned}
\gamma_{c\in{\{r,g,b\}}}=F(\delta_c)=\sum_{j=1}^{n}{P_j(\delta_c)}; \\
where\quad P_{j}(\delta_c)=c_{j}\prod _{\begin{smallmatrix}k=1; k\neq j\end{smallmatrix}}^{n}{\frac {\delta_{c}-\delta_{ck}}{\delta_{cj}-\delta_{ck}}}.
\end{aligned}
\label{lagrange-poly-rgb}
\end{equation}
}

\begin{table}[h]
  \centering
  \caption{Normalized $\delta_r$, $\delta_g$ and $\delta_b$ for different brightness levels of $RGB(1:1:1)$, using the LIFX bulb.}
    \begin{tabular}{rrrr}
    \toprule
    Composition $\langle r,g,b\rangle$ &  $\delta_r$    &  $\delta_g$    &  $\delta_b$ \\
    \midrule
    $\langle 0.2,0.2,0.2\rangle$    & 0.04052 & 0.013536 & 0.058512 \\
    $\langle 0.4,0.4,0.4\rangle$    & 0.149747 & 0.050023 & 0.209917 \\
    $\langle 0.6,0.6,0.6\rangle$    & 0.355869 & 0.115935 & 0.485289 \\
    $\langle 0.8,0.8,0.8\rangle$    & 0.66887 & 0.218776 & 0.906446 \\
    $\langle 1.0,1.0,1.0\rangle$    & 0.741883 & 0.242022 & 1 \\
    \bottomrule
    \end{tabular}
  \label{white-rgb-table}
\end{table}

{
\begin{equation}
\begin{aligned}
\gamma_{r} = 9.8923\delta_r^4 - 8.6910\delta_r^3 - 0.2037\delta_r^2 + 2.0864\delta_r \\+ 0.1163;\\
\gamma_{g} = 979.13\delta_g^4 - 304.98\delta_g^3 + 7.1778\delta_g^2 + 5.8836\delta_g \\+ 0.1198;\\
\gamma_{b} = 3.5353\delta_b^4 - 4.6515\delta_b^3 + 0.6081\delta_b^2 + 1.3907\delta_b \\+ 0.1174.
\end{aligned}
\label{lagrange-rgb}
\end{equation}
}

\section*{Appendix B - Exploratory Setups} 
\begin{figure}[h]
\centering
\begin{subfigure}{0.49\linewidth}
\centering
\includegraphics[width=\textwidth]{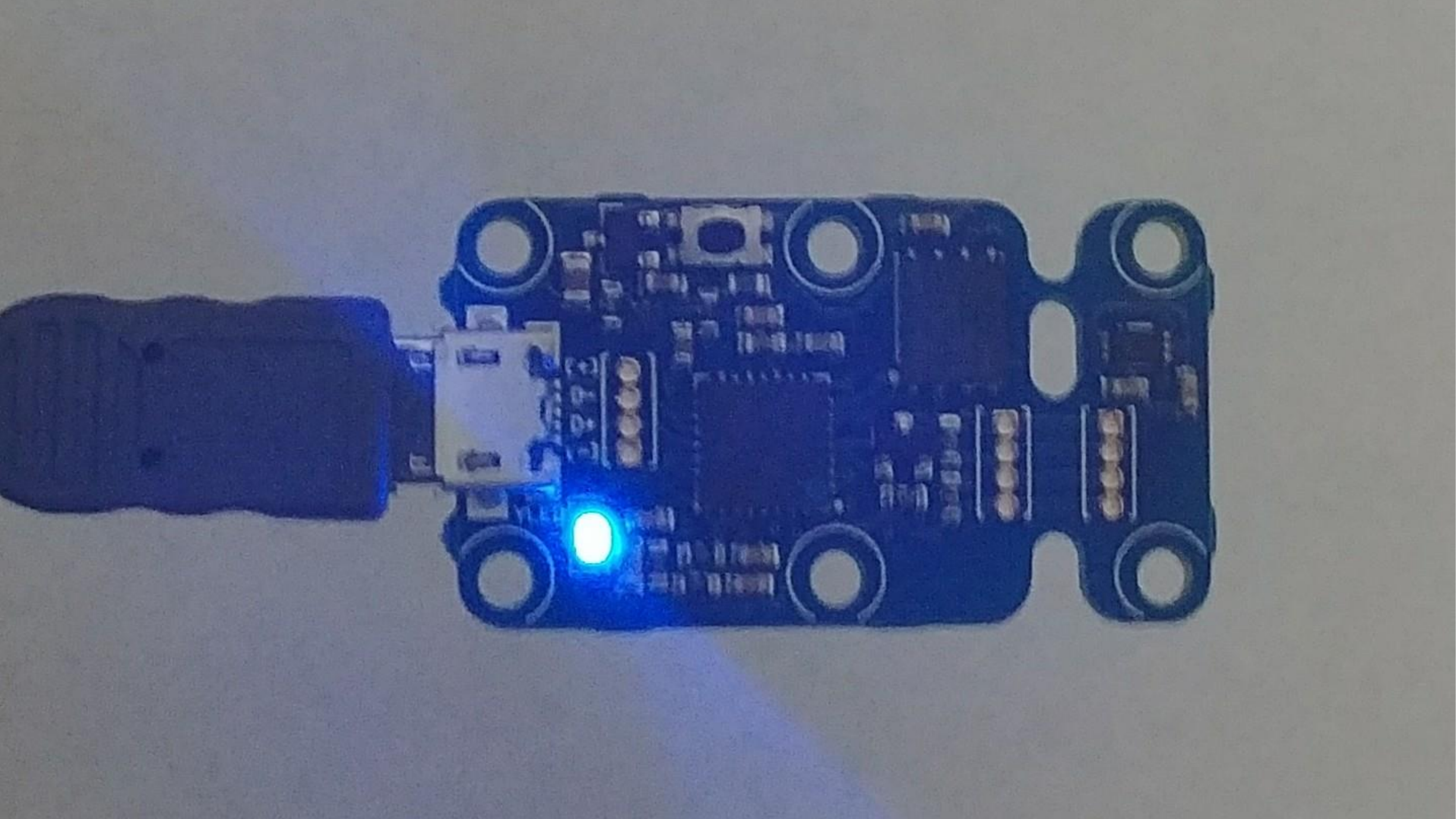}
\caption{}
\label{lightsensor}
\end{subfigure}
\hfill
\begin{subfigure}{0.49\linewidth}
\centering
\includegraphics[width=\textwidth]{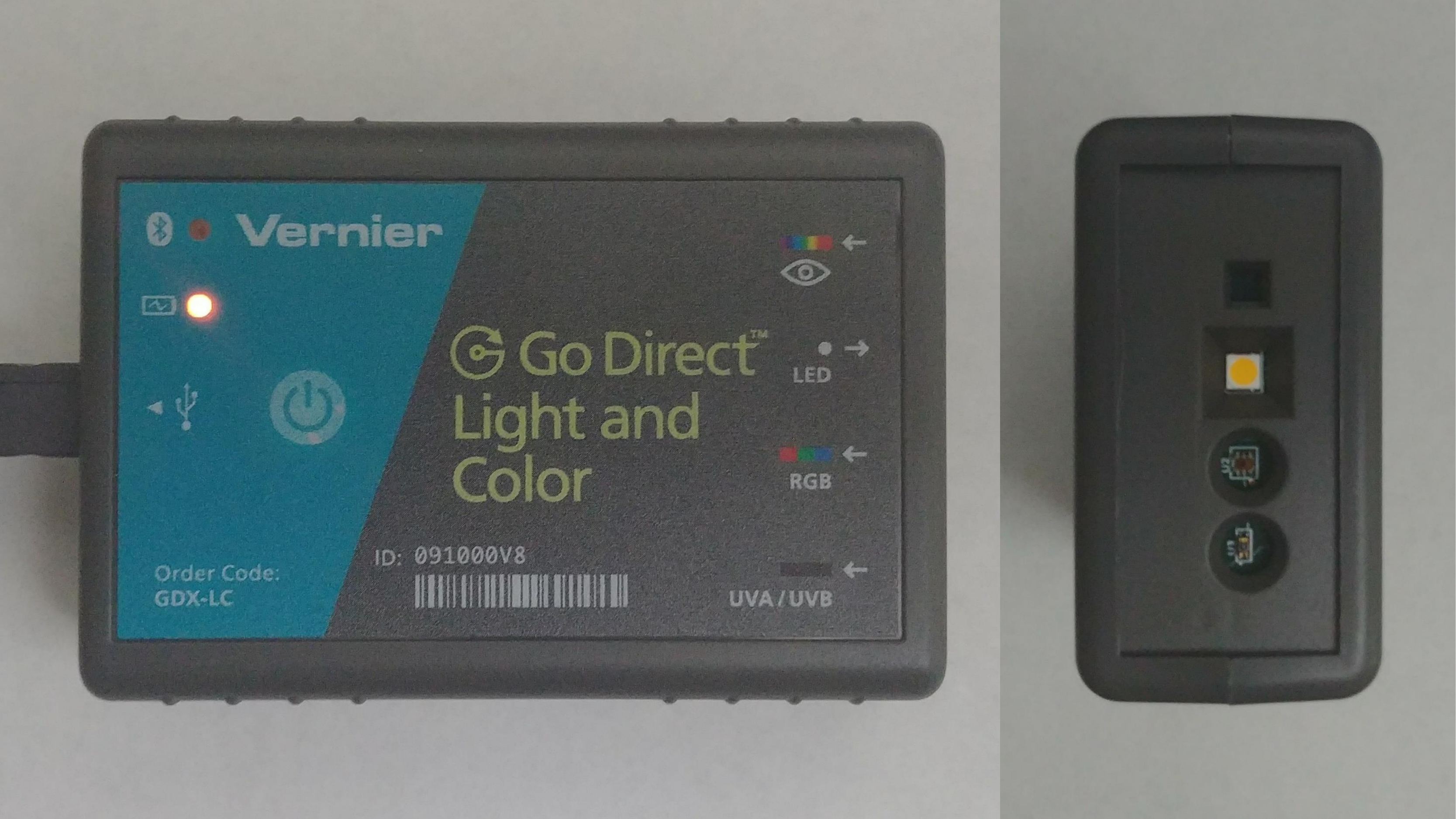}
\caption{}
\label{rgbsensor}
\end{subfigure}
\begin{subfigure}{0.35\linewidth}
\centering
\includegraphics[width=\textwidth]{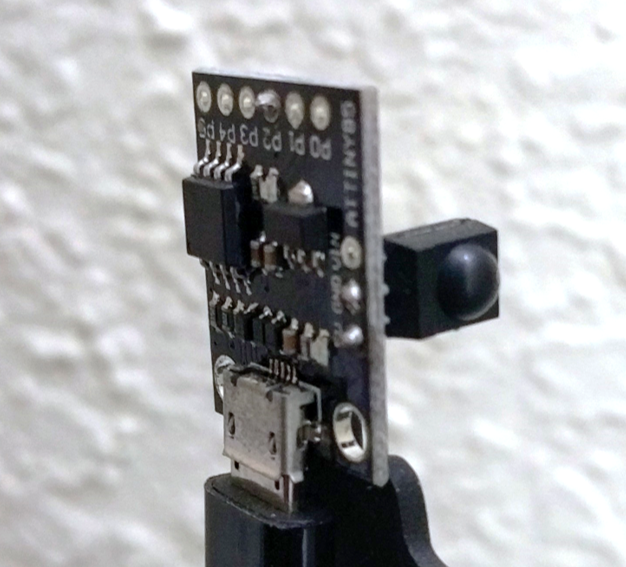}
\caption{}
\label{irsensor}
\end{subfigure}
\caption{(a) Yoctopuce V3 (BH1751FVI) luminance sensor; (b) Vernier GDX-LC RGB sensor; (c) TSOP48 infrared sensor soldered on an ATtiny85 Arduino board.}
\label{exploratory-sensors}
\end{figure}

\noindent
\textbf{Audio-visualization.} In the exploratory setup we used a\linebreak Yoctopuce V3 (BH1751FVI) luminance sensor for measuring light intensity (Figure \ref{lightsensor}). A LIFX A19 Gen3 bulb was used for audio-visualization. A 20 $mm$ 12x telephoto lens was used to focus light on the sensor.

\noindent
\textbf{Video-visualization.} In the exploratory setup we used a\linebreak Vernier GDX-LC RGB sensor to measure RGB composition of observed light (Figure \ref{rgbsensor}). A LIFX A19 Gen3 bulb was used for video-visualization. A 20 $mm$ 12x telephoto lens was used to focus light on the sensor.

\noindent
\textbf{Infrared exfiltration.} In the exploratory setup we used a\linebreak TSOP48 infrared sensor soldered on an ATtiny85 Arduino board for measuring 950 $nm$ infrared light (Figure \ref{irsensor}). A LIFX A19+ Gen3 bulb was used as a $M$-ary ASK transmitter, using the infrared spectrum. A 80 $mm$ 45-225x telescope was used to focus light on the sensor.

%

\end{document}